\documentclass[fleqn,usenatbib]{mnras}

\usepackage{newtxtext,newtxmath}

\usepackage[T1]{fontenc}

\DeclareRobustCommand{\VAN}[3]{#2}
\let\VANthebibliography\thebibliography
\def\thebibliography{\DeclareRobustCommand{\VAN}[3]{##3}\VANthebibliography}

\usepackage{graphicx}	
\usepackage{amsmath}	
\usepackage{siunitx}
\usepackage{pifont}
\newcommand{\angstrom}{\text{\normalfont\AA}}

\title[]{RAMSES-RTZ: Non-Equilibrium Metal Chemistry and Cooling Coupled to On-The-Fly Radiation Hydrodynamics}

\author[H. Katz] {Harley Katz$^{1}$\thanks{E-mail:
  \href{mailto:harley.katz@physics.ox.ac.uk}{harley.katz@physics.ox.ac.uk}}
  \\
  $^1$Sub-department of Astrophysics, University of Oxford, Keble Road, Oxford, OX1 3RH
  }

\date{Accepted XXX. Received YYY; in original form ZZZ}

\pubyear{2021}

\begin{document}
\label{firstpage}
\pagerange{\pageref{firstpage}--\pageref{lastpage}}
\maketitle

\begin{abstract}
Emission and absorption lines from elements heavier than helium (metals) represent one of our strongest probes of galaxy formation physics across nearly all redshifts accessible to observations. The vast majority of simulations that model these metal lines often assume either collisional or photoionisation equilibrium, or a combination of the two. For the few simulations that have relaxed these assumptions, a redshift-dependent meta-galactic UV background or fixed spectrum is often used in the non-equilibrium photoionisation calculation, which is unlikely to be accurate in the interstellar medium where the gas can self-shield as well as in the high-redshift circumgalactic medium where locally emitted radiation may dominate over the UV background. In this work, we relax this final assumption by coupling the ionisation states of individual metals to the radiation hydrodynamics solver present in {\small RAMSES-RT}. Our chemical network follows radiative recombination, dielectronic recombination, collisional ionisation, photoionisation, and charge transfer and we use the ionisation states to compute non-equilibrium optically-thin metal-line cooling. The fiducial model solves for the ionisation states of C, N, O, Mg, Si, S, Fe, and Ne in addition to H, He, and H$_2$, but can be easily extended for other ions. We provide interfaces to two different ODE solvers that are competitive in both speed and accuracy. The code has been benchmarked across a variety of gas conditions to reproduce results from {\small CLOUDY} when equilibrium is reached. We show an example isolated galaxy simulation with on-the-fly radiative transfer that demonstrates the utility of our code for translating between simulations and observations without the use of idealised photoionisation models. 
\end{abstract}




\section{Introduction}
Understanding both the distribution and ionisation states of metals throughout the Universe is key for a complete model of galaxy formation \citep[e.g.][]{Maiolino2019}. 

From an observational perspective, emission and absorption lines of elements heavier than hydrogen and helium represent one of the primary tools for probing the physics that is governing galaxy formation in a variety of astrophysical environments. For example, strong line diagnostics are widely used to determine interstellar medium (ISM) conditions \citep[e.g.][]{Baldwin1981,Kewley2019} or probe star formation \citep[e.g.][]{Kewley2004,DeLooze2014}. Metal absorption lines can be tracers of numerous physical processes such as the baryon cycle and circumgalactic medium (CGM) physics \citep[e.g.][]{Tumlinson2017,Ford2014}, kinematics of gas around galaxies \citep[e.g.][]{Steidel2010}, Lyman Continuum escape \citep[e.g.][]{Chisholm2020,Valentin2021}, as well as represent unique tests of galactic feedback \citep[e.g.][]{Keating2016}.

From a theoretical perspective, understanding the individual ionisation states of metals is paramount for accurately modelling gas cooling in simulations \cite[e.g.][]{Wiersma2009}. Furthermore, the exact abundances of individual metals can be used to constrain formation scenarios of different galactic components \citep{Kobayashi2006,Agertz2021}.

Various state-of-the-art numerical simulations employ very different approaches to modelling metal enrichment, cooling, and metal ionisation in simulations. It is now common practice to model the enrichment of individual heavy elements via various processes such as Type II supernovae (SNe), Type Ia SNe, and AGB winds \citep[e.g.][]{Schaye2015,Vogelsberger2014,Dubois2014}. However, there exists a wide diversity of models for metal line cooling. For low-density plasmas, a common assumption is that the gas exists in collisional ionisation equilibrium (CIE), in which case historical cooling functions \citep[e.g.][]{Sutherland1993} suffice as a representation of the gas thermodynamics. However, the Universe is permeated by a UV background which breaks the assumption of CIE and thus various cooling functions have been developed that evolve with redshift in response to the background radiation \citep[e.g.][]{Shen2010,Smith2017}. More detailed models apply cooling functions for individual metal ions, once again under the assumption of either CIE \citep[e.g.][]{Gnat2007} or in the presence of a UV background \citep[e.g.][]{Wiersma2009}, and more recently including the effects of a UV background, local stellar radiation, and self-shielding \citep{Ploeckinger2020}. Since the cooling rate is very sensitive to the shape and amplitude of the radiation field, assuming a UV background is a poor approximation in both self-shielded regions and in regimes where local sources dominate over the background. Modern radiation transfer simulations are now beginning to adopt cooling functions that explicitly account for local sources, but still under the assumption of equilibrium \citep[e.g.][]{Gnedin2012}.

When the individual ionisation states of metals are followed, metal line cooling can be computed exactly by solving for the level populations of electrons within the atom. \cite{Glover2007} present a model applicable for low-metallicity environments that relies on only O, C, and Si at low temperatures. \cite{Maio2007} follow a similar approach with the addition of Fe lines. \cite{Richings2014} present one of the most complete models for cooling in the ISM and find that the importance of individual lines is very sensitive to the shape and strength of the impinging radiation field. Finally, \cite{Ziegler2018} show that at higher temperatures (i.e., $T>10^4$K), typical assumptions that level populations can be accurately modelled by, for example, two, three, or five levels break down and significantly more levels need to be followed to properly model gas cooling.   

Due to its importance, tracking non-equilibrium metal chemistry is becoming more common in simulations \citep[e.g.][]{Oppenheimer2013,Richings2014,Bovino2016,Ziegler2018,Sarkar2021}, especially as open-sourced chemistry libraries continue to be developed \citep[e.g.][]{Grassi2014}. However, few of these works couple the chemistry to the inhomogeneous radiation fields that can only be captured by on-the-fly radiation hydrodynamics. Notable exceptions are \cite{Baczynski2015} who use the {\small FLASH} code but only follow a small network of C{\small II}, and CO, and \cite{Sarkar2021} who use the {\small PLUTO} code to model a very detailed chemical network coupled with radiative transfer (RT) using the short characteristics method. Unfortunately, this method of RT is not well suited for cosmological simulations because the computational load scales approximately linearly with the product of the number of grid cells and the number of sources. Most similar to the method presented in this work is \cite{Lupi2020} who couple a very small network of O, C, and Si to the M1 radiative transfer in {\small GIZMO} to predict far infrared emission line luminosities at high-redshift.

In this work, we build upon these previous approaches by coupling non-equilibrium metal chemistry and cooling to the radiation hydrodynamics code {\small RAMSES-RT} \citep{Teyssier2002,Rosdahl2013,Rosdahl2015}. Our goals are to more accurately model gas cooling as well as emission and absorption lines from individual metals efficiently enough for use in cosmological simulations, thus significantly improving on the approach we have recently used to compare simulations with observations \citep{Katz2017,Katz2019,Katz2021}. 

{\small RAMSES-RT} is particularly well suited for cosmological simulations because it employs the M1 method for RT \citep{Levermore1984}, which does not scale with the number of sources in the computational volume and hence can be used for large scale simulations of both individual environments \citep[e.g.][]{Trebitsch2021} as well as full box simulations \citep[e.g.][]{Rosdahl2018}. Furthermore, the code makes use of the variable-speed-of-light approximation \citep{Katz2017}, which is particularly useful for reionization simulations and is able to run with both RT and ideal magnetohydrodynamics (MHD) simultaneously \citep[e.g.][]{Katz2021,Kimm2021}, which are important for studies of the CGM \citep[e.g.][]{Mitchell2021,vdv2021,Buie2022}. As the importance of following non-equilibrium metal ionisation states and cooling is well established in the literature \citep[e.g.][]{Oppenheimer2013}, we present in this work our implementation in {\small RAMSES-RT}.

This work is organised as follows. In Section~\ref{methods}, we describe our methods for implementing non-equilibrium metal chemistry and cooling. In Section~\ref{bench}, we present a series of benchmarks where we have tested various physical processes to show that the code reaches the correct equilibrium abundances as predicted by {\small CLOUDY}. In Section~\ref{otf}, we show an example simulation of an isolated galaxy, highlighting how {\small RAMSES-RTZ} simulations can be used to predict emission lines and strong-line diagnostics without the use of idealised, equilibrium photoionisation models. Finally, in Sections~\ref{cavs} and \ref{dac}, we present our caveats and conclusions.

\section{Method}
We first describe the methods for evolving metal ionisation states in {\small RAMSES-RT} and then the non-equilibrium cooling routines.

\begin{table}
    \centering
    \caption{List of symbols used in this work and their meanings.}
    \begin{tabular}{ll}
        Symbol & Meaning \\
        \hline
        $\alpha$ & Total recombination rate coefficient (radiative + dielectronic)   \\
        $\beta$ & Collisional ionisation rate coefficient with electrons \\
        $\xi$ & Charge exchange recombination coefficient \\
        $\chi$ & Charge exchange ionisation coefficient \\
        $\Gamma$ & Photoionisation rate \\
    \end{tabular}
    \label{tab:my_label}
\end{table}

\label{methods}
\subsection{Chemistry}
The time evolution of the ionisation states of individual metals are dependent on numerous processes including radiative recombination, dielectronic recombination, collisional ionisation, photoionisation, Auger ionisation and charge exchange. In general the creation of ion, $n_i$, (excluding Auger ionisation) can be computed as
\begin{equation}
\label{eq1}
\begin{split}
    \frac{dn_i}{dt} = n_{i+1}\alpha_{i+1}n_e + \left(\sum_k [n_{i+1}\xi_{i+1}+n_{i-1}\chi_{i-1}]n_k\right) \\ + n_{i-1}(\beta_{i-1}n_e+\Gamma_{i-1}),
\end{split}
\end{equation}
where $\alpha_{i+1}$ is the total recombination rate (radiative and dielectronic) of the more ionised state, $\xi_{i+1}$ and $\chi_{i-1}$ are the charge exchange recombination and ionisation coefficients for the more ionised and less ionised states, respectively for each charge exchange partner, $k$, $\beta_{i-1}$ is the collisional ionisation rate coefficient with electrons for the less ionised state, and $\Gamma_{i-1}$ is the photoionisation rate of the less ionised state. Note that $\beta$ can be generalised for collisions with species other than electrons. The destruction of ion $n_i$ (again excluding Auger ionisation) follows:
\begin{equation}
\label{eq2}
    \frac{dn_i}{dt} = -n_{i}\alpha_{i}n_e - \left(\sum_k [n_{i}\xi_{i}+n_{i}\chi_{i}]n_k \right) - n_{i}(\beta_{i}n_e+\Gamma_{i}),
\end{equation}
where the symbols represent the same processes as above; however, the subscripts have been updated to represent the fact that they now apply to the current ionisation state. In general, all of the coefficients are sensitive to both the temperature and often the density of the gas and can be generalised is the presence of other physical processes (e.g., cosmic rays). Combining Equations~\ref{eq1} and \ref{eq2} allows us to solve for the time evolution of each ion.  We highlight that there is a coupling between different species both through the electron density and through charge exchange reactions. 

We compile values for these coefficients from a variety of sources and similar to \cite{Oppenheimer2013}, we have designed our models to follow closely to the data used by {\small CLOUDY} \citep{Ferland2017}. Most radiative and dielectronic recombination rates are taken from \cite{Badnell2006} and \cite{Badnell2003}, respectively. These sources do not provide rates for the low ionisation states of heavier elements and thus we resort to a mixture of radiative and dielectronic recombination rates\footnote{As compiled at https://www.pa.uky.edu/$\sim$verner/rec.html.} from \cite{Aldrovani1973,Shull1982,Arnaud1985}. For Fe, radiative and dielectronic rates are taken from \cite{Arnaud1992} when not available in \cite{Badnell2006} and \cite{Badnell2003}. We do not include dielectronic recombination suppression \citep[e.g.][]{Nikolic2013,Nikolic2018} as the effect is primarily important at electron densities $\gtrsim10^5{\rm cm^{-3}}$, which are higher than those we intend to probe with the code. Fitting functions for collisional ionisation rates are taken from \cite{Voronov1997}. 

To calculate the photoionisation rates, we compute photon advection on-the-fly with the code. Data for atomic cross sections as a function of frequency is from \cite{Verner1996}. The vast majority of charge exchange recombination and ionisation rates for ions with hydrogen are taken from \cite{Kingdon1996}. Certain rates, for example those of low ionisation states of oxygen, have been updated in {\small CLOUDY}. We have updated the rate equations for oxygen charge exchange reactions to match those from the most recent version (v17.02) of the {\small CLOUDY} source code \citep{Stancil1999,Barragan2006}. For all others, we rely on \cite{Kingdon1996}\footnote{Note that certain charge exchange rates for Fe and S have also been updated in {\small CLOUDY} and small discrepancies may be found between {\small RAMSES-RTZ} and {\small CLOUDY} for this reason.}. We have made the conscious decision not to include charge exchange reactions between helium and metals. Including such reactions is only expected to change the result by a mild amount because the hydrogen number density dominates\footnote{Because of their different ionisation energies and recombination rates, depending on the radiation source population and gas temperature, there may be scenarios where He{\small I} is present when H{\small I} is not. This can also happen for He{\small II} and H{\small II}. Furthermore, if stellar feedback (i.e., winds or SNe) return more He than H, our primordial number density argument may not hold. In these scenarios, it is possible that charge exchange with helium temporarily dominates and such situations will not be accurately modelled by our code.} over helium. Furthermore, reaction rates are not available between all metal ions and helium which could bias our results. Moreover, many of the fitting functions for charge exchange reactions between helium and metals (e.g., such as those used by {\small CLOUDY}) are not all well behaved at all temperatures. Similarly we do not include charge exchange reactions between different metal species since the number densities are so low that the impact on our results will be small. Finally our model excludes Auger ionisation which is expected to only have a minor impact on the ionisation states \citep{Oppenheimer2013}. We stress that updating our network to include the processes that we exclude in our fiducial model is trivial. Our selection of physical processes is based primarily upon considerations of computation speed and the accuracy loss of excluding the physical process.

Our method for integrating the ionisation states of individual species over each simulation time-step differs significantly from other works. In principle, one should integrate the temperature, ionising photon number densities, and species ionisation states simultaneously and implicitly. This is trivial to set up, but slow in practice because the reaction coefficients are often temperature (and sometimes density) dependent and the couplings between many of the reactions means that achieving converged results requires the integrator to take numerous sub-steps, which can be extremely computationally expensive. For this reason, we have applied a series of operator splits in order to make our code more efficient. Our method for integrating the ionisation states is as follows.

The temperature, photon number densities, H, H$_2$, and He updates are nearly identical to those in \cite{Rosdahl2013,Katz2017} with the following exceptions. We have now included charge exchange reactions between H and He as well as H and the metals. The electron density used for the H, H$_2$, and He update now accounts for the presence of metals. Furthermore, we explicitly subtract off the metal mass from the gas mass in the cell when updating H and He (which are often assumed to have fixed mass fractions of 76\% and 24\%) so that the code is self-consistent. Finally, we also include the metals as sinks for photons. Besides the addition of charge transfer between H and He, these further updates are minor and only included for self-consistency. We also treat metal line cooling differently from what is standard in {\small RAMSES}, as described in Section~\ref{sec:cooling}, but the order in which the temperature is updated with respect to the other integrations is consistent with {\small RAMSES-RT}.

\begin{figure}
\centerline{\includegraphics[scale=1,trim={0 0.7cm 0cm 1.0cm},clip]{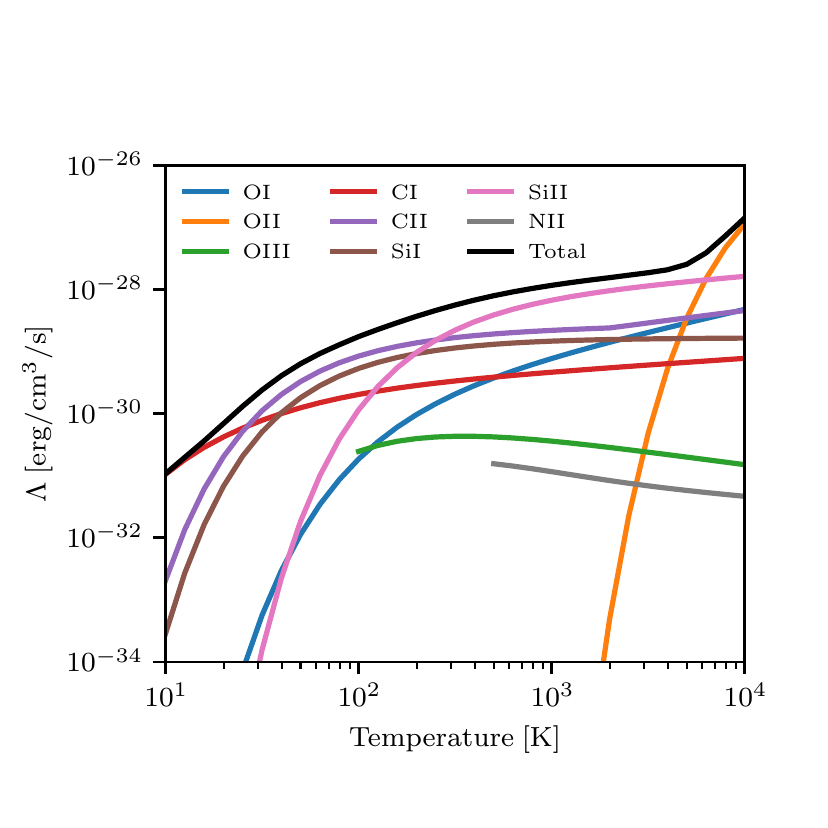}}
\caption{Cooling rates as a function of temperature assuming a neutral hydrogen density of 1cm$^{-3}$, an electron density of $10^{-4}{\rm cm^{-3}}$, a proton density $10^{-4}{\rm cm^{-3}}$, a metal ion density of $10^{-6}{\rm cm^{-3}}$ and no H$_2$ (similar to \protect\citealt{Grassi2014}). Cooling rates have been computed under the assumption of a $z=0$ CMB. Note that cooling from O{\small III} and N{\small II} are not computed at $T<100$K and $T<500$K, respectively, hence the lines are truncated at these values. }
\label{cooling_z0}
\end{figure}

After the temperature, photon number densities, H, H$_2$, and He have been updated, we then update the metal ionisation states. We provide two interchangeable methods for doing so. First, we have included an interface to the {\small DVODE}\footnote{https://www.radford.edu/$\sim$thompson/vodef90web/} BDF solver which is called for each species individually. Because we have applied an operator split for temperature, we can compute the Jacobian analytically and provide it directly to the solver. This is our fiducial method and it is expected to be the most accurate; however, the accuracy comes with a significant computational cost. Because of this, we provide a second method which is significantly faster at the potential expense of accuracy. The method is very similar to the method used in {\small RAMSES-RT} which takes inspiration from \cite{Anninos1997}. We cycle through the ionisation states and update them semi-implicitly using forward-in-time values for all previous (i.e., lower) ionisation states, H, He, $e^-$, temperature, and photon number density, but backwards-in-time values for all ionisation states that have yet to be updated (i.e., higher). More specifically, we solve for the ionisation fractions of each ion ($x_i$) from the ground state first such that
\begin{equation}
    x_i^{t+\Delta t}=\frac{x_i^{t}+C\Delta t}{1+D\Delta t},
\end{equation}
where $C$ is equivalent to the right-hand side of Equation~\ref{eq1}, D is equivalent to the right-hand side of Equation~\ref{eq2}, and $\Delta t$ is the simulation time-step. As argued by \cite{Anninos1997,Rosdahl2013}, this is partially implicit because it relies on already updated values (i.e., temperature H, He, $e^-$, photon density, and all lower ionisation states) but it still depends on the un-updated values of the current ionisation fraction and all higher ionisation states. When combined with a 10\% rule\footnote{The 10\% rule prevents individual parameters (i.e., ionisation fractions, temperature, photon densities, etc.) from changing by more than 10\% during any cooling sub-cycle. If the 10\% rule is violated, the cooling/chemistry step is repeated with a smaller $\Delta t$ until the condition is satisfied.} as outlined in \cite{Rosdahl2013} we empirically find that this method is very stable and significantly faster than the {\small DVODE} BDF solver. The key component that makes this method faster is that in principle, only one call is needed for each simulation time step. Because of the RT Courant condition, the simulation time step is often very short compared to the sound crossing time or the timescales related to the chemistry. Hence, sub-steps are unnecessary except when triggered by the 10\% rule. Since one step of the fast method is faster than the accurate solver, the fast method provides an enticing alternative to the more standard method outlined above.  

For the remainder of this work, we refer to the initial method as the ``accurate'' method and the second method as the ``fast'' method.

\subsection{Cooling \& Line Emission}
\label{sec:cooling}
Having access to individual metal ionisation states provides the opportunity to calculate the gas cooling rates in a non-equilibrium manner. When there is no external radiation field or the radiation is easily parametrisable (e.g., a redshift-dependent UV background), cooling rates for individual ions can be tabulated \citep[e.g.][]{Gnat2007,Vasiliev2011,Oppenheimer2013,Richings2014}. Recent work has extended such tabulations to account for interstellar radiation fields, self-shielding, and cosmic rays \citep{Ploeckinger2020}. However, when both the shape and intensity of the radiation field strongly vary, as is the case in radiation hydrodynamics simulations, cooling must be computed by solving for the level populations of a given ionisation state \citep[e.g.][]{Grassi2014,Ziegler2018}. The additional computational cost of solving for the level populations on-the-fly can be significant, especially when the ion is characterised by more than three levels. For this reason, we have taken a simplified approach (compared to e.g. \citealt{Ziegler2018}) as follows.

\begin{figure*}
\centerline{\includegraphics[scale=1,trim={0 0 0 0},clip]{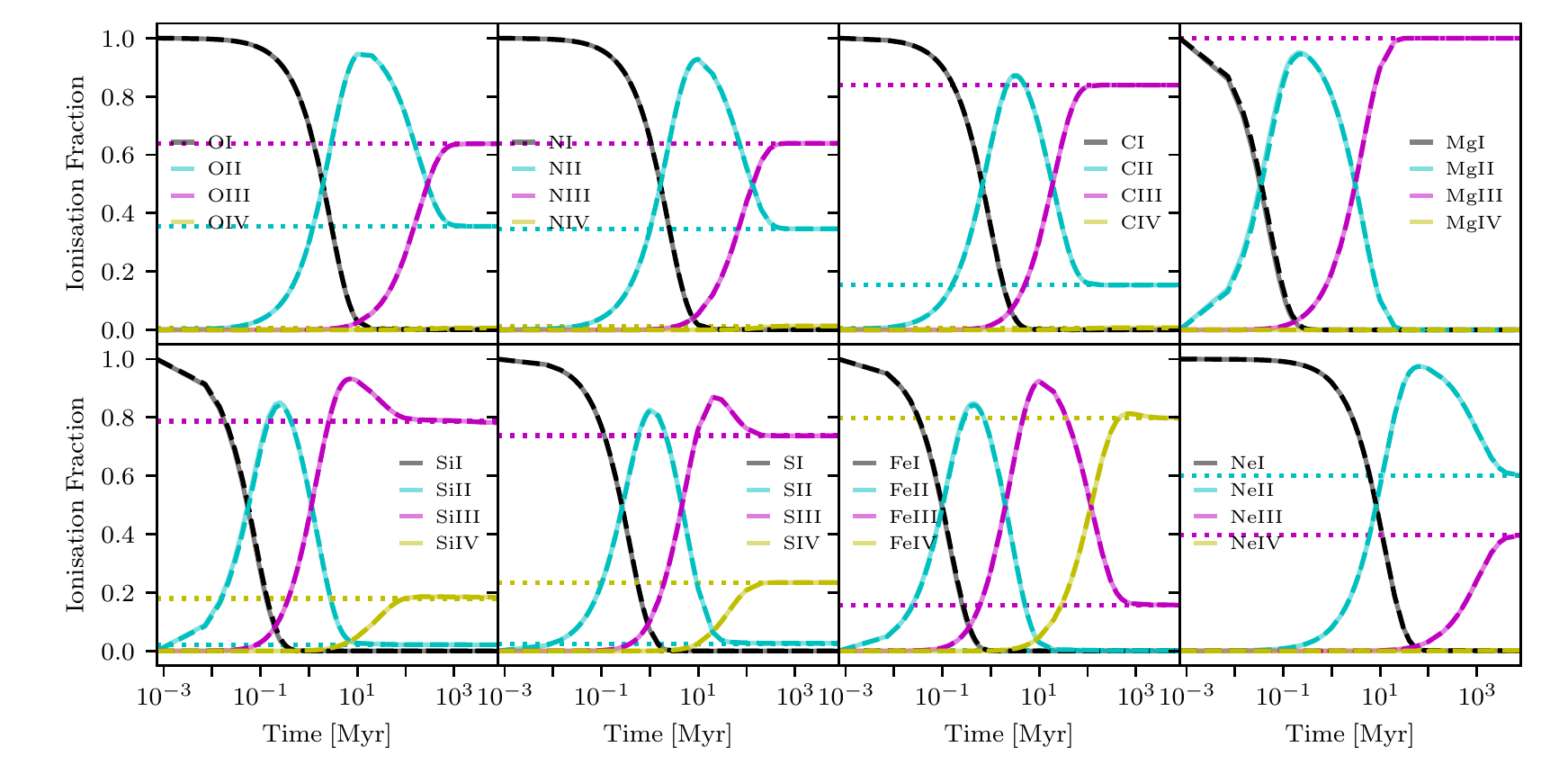}}
\caption{Ionisation fractions as a function of time for the first four ionisation states of O, N, C, Mg, Si, S, Fe, and Ne assuming only collisional ionisation. All elements are initialised in their neutral state with the gas density and pressure set to $10^{-5}{\rm cm^{-3}}$ and $P/k_{\rm B}=1\ {\rm cm^{-3}}$K, respectively. The solid and dashed lines show the results from the tests when using the accurate and fast solvers, respectively. The dotted lines represent the equilibrium abundances measured with {\small CLOUDY}. In all cases, the accurate and fast solvers agree with each other and the simulations reproduce the equilibrium abundances from {\small CLOUDY} when equilibrium has been reached. }
\label{coll_ion_test_0}
\end{figure*}

Atomic cooling processes of H and He are computed analytically in a non-equilibrium manner as is already done in {\small RAMSES-RT} \citep{Rosdahl2013}. For metal line cooling at $T>10^4{\rm K}$, we rely on the ion-by-ion cooling tables provided by \cite{Oppenheimer2013}. If the ion is not followed by the code, we assume collisional ionisation equilibrium to compute the abundance. The abundance of untracked ions is then rescaled so that the total ion abundances for tracked and untracked ions sums to one for each element (see e.g. \citealt{Gray2015}). The code can be generalised for other assumptions (e.g., photoionization equilibrium with a UV background or an alternative radiation field, etc.). For $T\leq10^4{\rm K}$, metal enriched gas cooling is dominated by H$_2$ and metallic fine-structure emission lines \citep[e.g.][]{Hollenbach1989}. The non-equilibrium H$_2$ abundance and cooling is accounted for as in \cite{Katz2017} and the new aspect of this work in the context of the {\small RAMSES} code is that we explicitly calculate the fine structure emission from O{\small I}, O{\small II}, O{\small III}, C{\small I}, C{\small II}, Si{\small I}, Si{\small II}, N{\small II}, Fe{\small I}, Fe{\small II}, S{\small I} and Ne{\small II}. 

At low metallicity, \cite{Glover2007} argue that the cooling rate is dominated by O{\small I}, C{\small I}, C{\small II}, Si{\small I}, and Si{\small II} \citep[see also][]{Wolfire2003}. Thus these lines constitute our fiducial model. \cite{Maio2007} further include Fe{\small II}; however, depending on the epoch, cooling from Fe may be unnecessary. For example, the enrichment timescale for Fe is much longer O and C \citep[e.g.][]{Maiolino2019}, hence it is less important for the early Universe. Furthermore, \cite{Richings2014} showed Fe cooling is significantly less important when assuming realistic dust depletion factors \citep{Jenkins2009}. Thus cooling lines should be carefully chosen.

It is trivial to update {\small RAMSES-RTZ} to include lines from any ion; however, for a given simulation one we recommend balancing computation time and accuracy when choosing lines to include. The additional fine-structure lines that are available beyond the \cite{Glover2007} model are motivated by those that may be important for cooling following \cite{Richings2014} and those that are important for high-redshift observations (e.g. [C{\small II}] and [O{\small III}]\footnote{Note that [O{\small III}] 88$\mu$m emission, which can be exceptionally bright at $z>6$, originates in gas hotter than $10^4$K. Our cooling calculation for this line is applicable at higher temperatures and thus can be used to compute the emission line luminosity for this line; however, for cooling, we consider it only at $T\leq10^4{\rm K}$ since the ion-by-ion cooling tables we employ at hotter temperatures already account for the effect. The same holds true for N{\small II} cooling and emission.}).  
For computational efficiency, we make two simplifying approximations. First, we assume that cooling can be calculated in the optically thin limit which allows us to exclude effects from self-absorption. This is reasonable at gas densities approximately below the critical density of the transition. Second, we assume that O{\small I}, O{\small II}, O{\small III}, C{\small I}, Si{\small I}, N{\small II}, Fe{\small I}, Fe{\small II}, and S{\small I} can be modelled as three-level ions while C{\small II}, Si{\small II}, and Ne{\small II} require only two levels. This second assumption allows us to model the level populations with a simple set of analytical expressions whereas including more levels would likely necessitate the use of a linear equation solver (e.g., {\small LAPACK}).

Following \cite{Glover2007}, for O{\small I}, C{\small I}, C{\small II}, Si{\small I}, and Si{\small II} we include collisional rates for $e^-$, H, H$^+$, and H$_2$ (ortho and para assuming a 3:1 ratio), where available. All collisional rates, Einstein coefficients, and level population data can be found in their Table~5 and 6. For O{\small II} we include only collisions with $e^-$, with data consistent with \cite{Grassi2014}. Similarly, for O{\small III}, we include only collisions with $e^-$. Einstein coefficients are taken from \cite{Garstang1968} and level population information is from \cite{Draine2011} (consistent with \citealt{Yang2020}). Furthermore, we have fit higher-order polynomials to the collision strength data from \cite{Storey2014} in order to calculate the collisional de-excitation rates with electrons. Einstein coefficients for N{\small II} are also from \cite{Garstang1968} and we have fit higher order polynomials to electron collision strength data from \cite{Tayal2011}. Finally, data for cooling from Fe{\small I}, Fe{\small II}, S{\small I}, and Ne{\small II} is taken from \cite{Hollenbach1989}. Note that we include the simulated emission coefficients in our model which allows us to take into account the impact of both the cosmic microwave background as well as the local radiation field \citep[e.g.][]{Glover2007}, if desired.

In Figure~\ref{cooling_z0}, we show an example set of cooling rates computed under the same conditions as listed in Figure~3 of \cite{Grassi2014}. Depending on the temperature, different fine structure lines dominate the cooling, consistent with expectations from \cite{Maio2007,Grassi2014}.

\begin{figure}
\centerline{\includegraphics[scale=1,trim={0 0 0 0},clip]{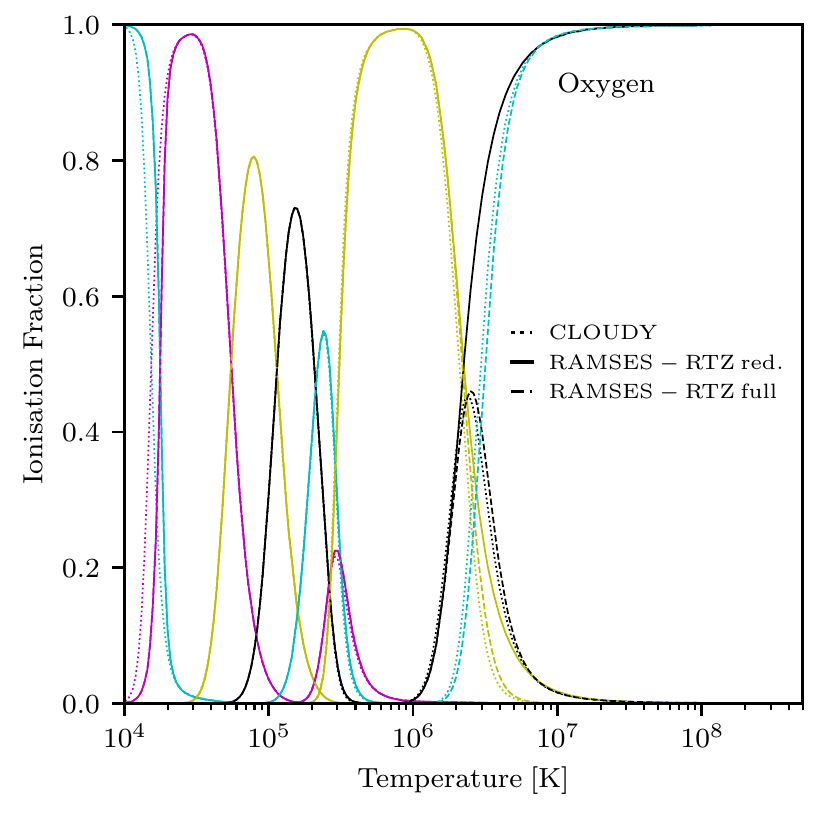}}
\centerline{\includegraphics[scale=1,trim={0 0 0 0},clip]{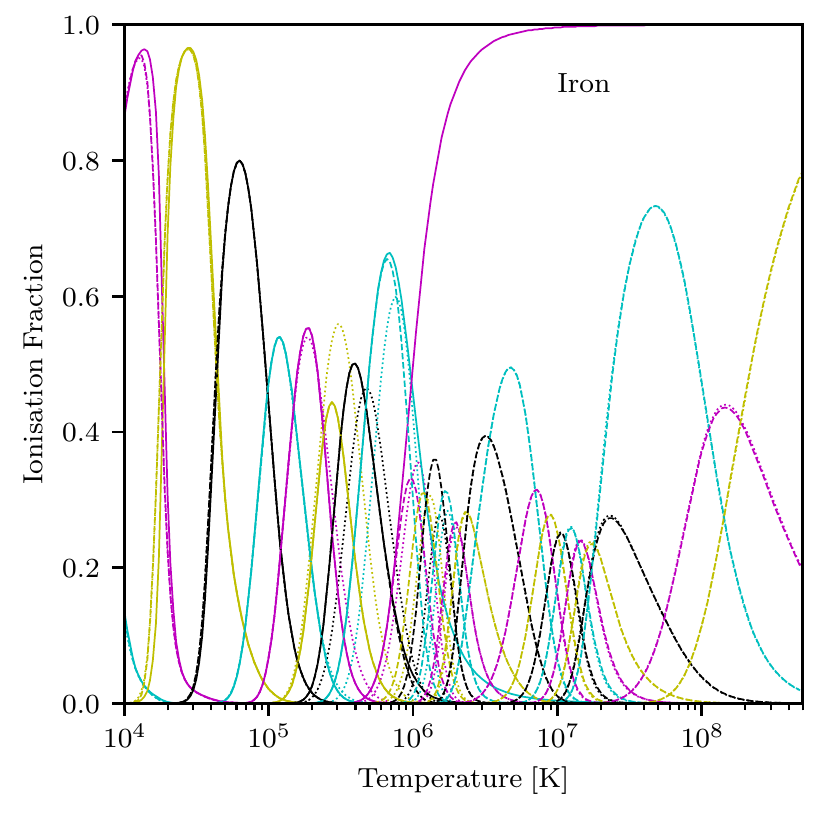}}
\caption{Oxygen (top) and iron (bottom) ionisation fractions as a function of temperature in collisional ionisation equilibrium at a constant density of $10^{-5}{\rm cm^{-3}}$. Solid and dashed lines show the results from {\small RAMSES-RTZ} using the reduced and full chemical networks, respectively. Dotted lines show the results from {\small CLOUDY}. The agreement between our code and {\small CLOUDY} is maintained across a wide range in temperature.}
\label{coll_ion_test_1}
\end{figure}

\begin{figure*}
\centerline{\includegraphics[scale=1,trim={0 0 0 0},clip]{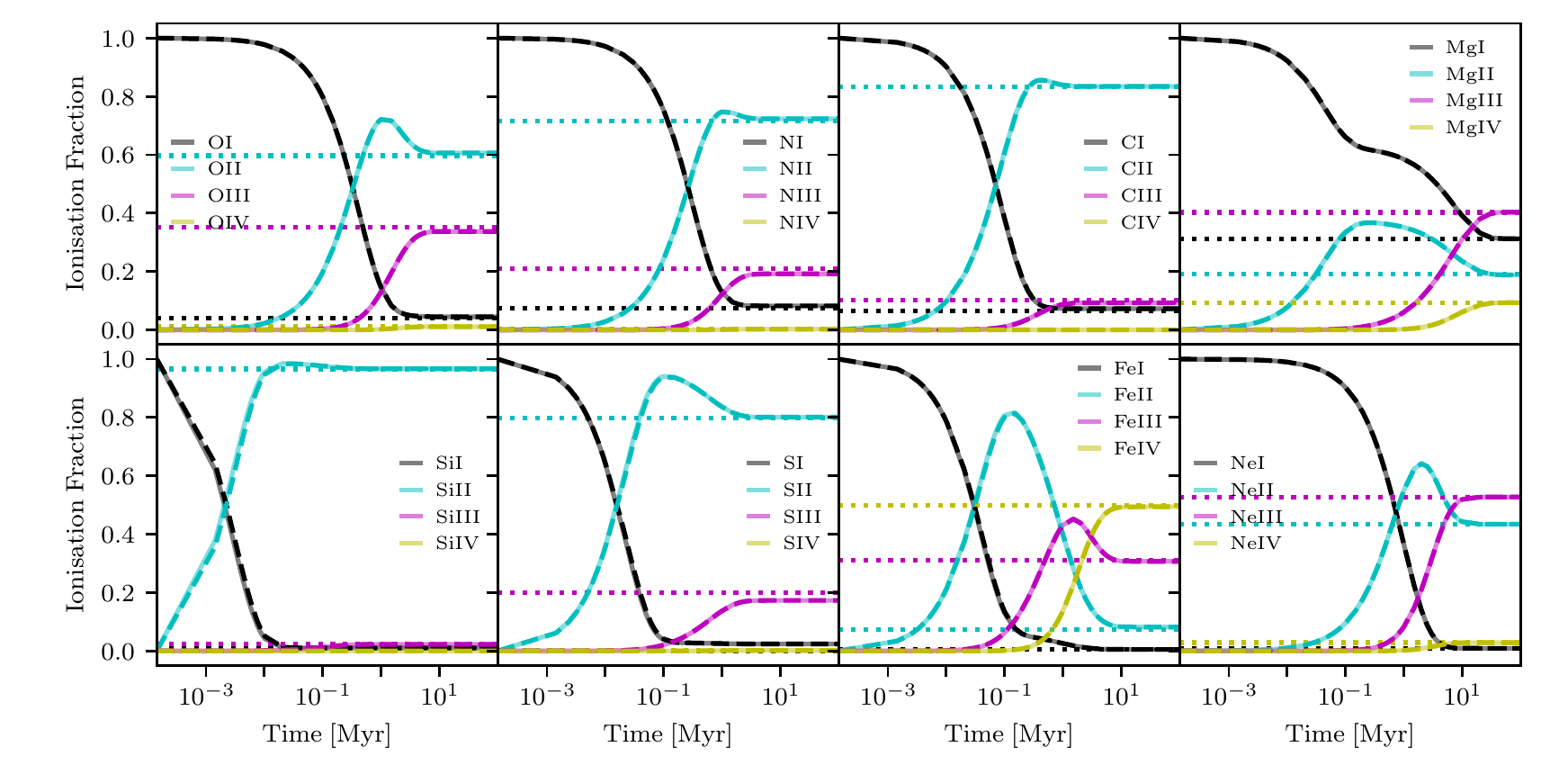}}
\caption{Ionisation fractions as a function of time for the first four ionisation states of O, N, C, Mg, Si, S, Fe, and Ne assuming only photoionisation from a \protect\cite{Haardt2012} UV background at $z=0$. Note that collisional ionisation has been turned off for this test. All elements are initialised in their neutral state with the gas density and pressure set to $10^{-2}{\rm cm^{-3}}$ and $P/k_{\rm B}=500\ {\rm cm^{-3}}$K, respectively. The solid and dashed lines show the results from the tests when using the accurate and fast solvers, respectively. The dotted lines represent the equilibrium abundances measured with {\small CLOUDY}. In all cases, the accurate and fast solvers agree with each other and the tests reproduce the equilibrium abundances from {\small CLOUDY} when equilibrium is reached. }
\label{phot_ion_test_0}
\end{figure*}

\section{Benchmarks}
\label{bench}
In this section we compare the results from a set of test simulations when they have reached equilibrium with the equilibrium solution from {\small CLOUDY}. The tests are conducted in the following way. We set up a $2\times2$ oct grid representing a 15~kpc box in {\small RAMSES} with outflow boundary conditions and assign the same initial density and pressure to all cells. In practice, these tests can be run on a single cell; however, we opt to ensure that all cells are evolved identically. Cooling and gravity are turned off so that the pressure and density are constant throughout the simulation. Hydrogen and helium are initialised in a neutral state with mass fractions of 76\% and 24\% of the primordial gas content, respectively. We assign solar metallicity to all cells with abundance ratios taken from \cite{Grevesse2010}. For simplicity, we run the tests including non-equilibrium ionisation states of oxygen (up to O{\small VIII}), nitrogen (up to N{\small VII}), carbon (up to C{\small VI}), magnesium (up to Mg{\small VI}), silicon (up to Si{\small X}), sulfur (up to S{\small X}), iron (up to Fe{\small X}), and neon (up to Ne{\small VIII}), where all elements are initialised in a neutral state. Other metals can be trivially added to the network (for example, the metal enrichment in our cosmological simulations also follows Ca, Katz et al. {\it in prep.}). We then turn individual processes on and off to ensure that our network is correct in all regimes where each process dominates.

For comparison with {\small CLOUDY}, we set up models, all at constant temperature, density, and metallicity that match the simulation. We use an open geometry slab with a thickness of 1~cm and a single zone, and iterate until convergence. For all calculations, Auger ionisation is turned off, molecules, dust, and induced processes are excluded, and we turn off dielectronic recombination suppression. Depending on the test, as further discussed below, we turn on collisional ionisation, photoionisation, and charge transfer as needed. While these steps ensure that the physics included in the {\small CLOUDY} models is about as close to the {\small RAMSES-RTZ} simulations as possible, we highlight the fact that there are minor differences in the reaction coefficients that prevent perfect agreement. 

Note that the {\small RAMSES-RTZ} simulations are run at constant pressure. Because heating and cooling processes are turned off, this sets a constant $T/\mu$, where $\mu$ is the mean molecular weight of the gas. To make a fair comparison with {\small CLOUDY}, we print out the equilibrium temperature in the {\small RAMSES-RTZ} simulation set by the final ionisation fractions and use this as the input temperature for the {\small CLOUDY} simulations.

\subsection{Collisional Ionisation}
In the first test of collisional ionisation, we set the gas density to $10^{-5}{\rm cm^{-3}}$ and the pressure to $P/k_{\rm B}=1\ {\rm cm^{-3}}$K (which corresponds to an equilibrium temperature of $\sim61,000$K). The low density of the gas and the moderate temperature means that the equilibrium timescale for many of the ions is on the order of Gyrs. The temperature is set to a low value so that only the first few ionisation states of each metal are populated.

\begin{figure}
\centerline{\includegraphics[scale=1,trim={0 0 0 0},clip]{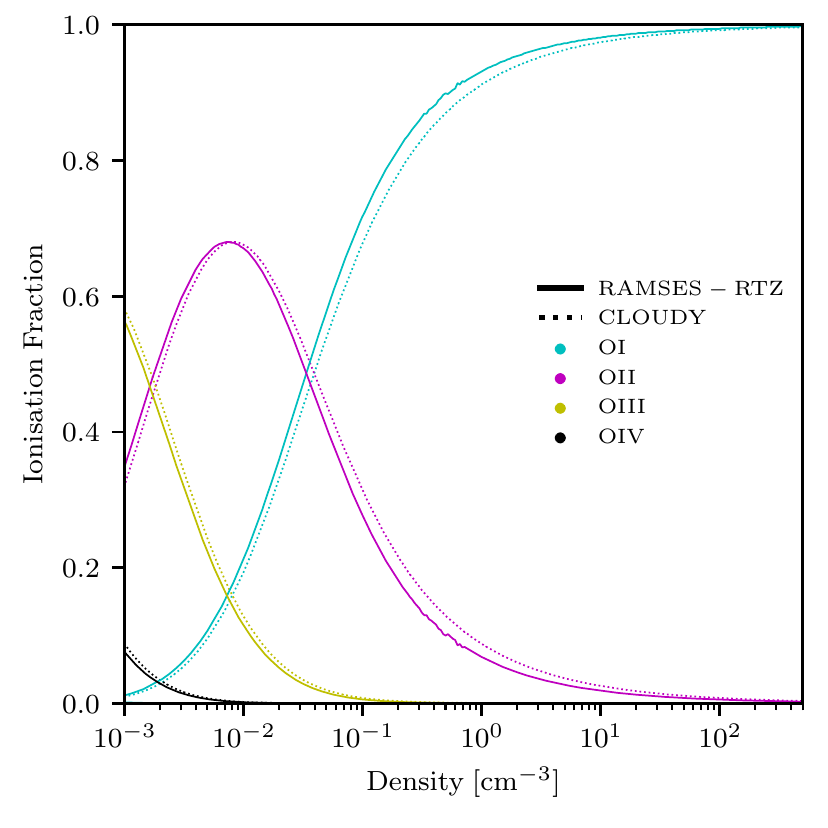}}
\caption{Oxygen ionisation fractions as a function of density in  photoionisation equilibrium assuming a $z=0$ \protect\cite{Haardt2012} UV background at a constant temperature of $T=5\times10^{4}{\rm K}$. Dotted lines show the results from {\small CLOUDY} while solid lines represent the results from our test. The agreement between our code and {\small CLOUDY} is maintained across a wide range in density.}
\label{phot_ion_test_1}
\end{figure}

In Figure~\ref{coll_ion_test_0}, we show the time evolution of the ionisation fraction of the first four ionisation states of O, N, C, Mg, Si, S, Fe, and Ne for our simulation (solid lines) compared to the equilibrium values obtained with {\small CLOUDY}\footnote{Note that the collisional ionisation rates used by {\small CLOUDY} are a hybrid of \cite{Voronov1997} and those from {\small CHIANTI} \citep{Dere2019}. In order to obtain an agreement, we have used the {\small CLOUDY} command ``{\small set collisional ionization Dima}''. Furthermore, when the rate is uncertain, {\small CLOUDY} will assume the mean of rates for a similar ionisation state for other ions. To avoid this behaviour and match the rates between {\small RAMSES-RTZ} and {\small CLOUDY}, we have used the command {\small ``set dielectronic recombination mean off''}.} (horizontal dotted lines). The test has been evolved until equilibrium is reached for all ions. In all cases, the results from the test at the final output are in very good agreement with the values predicted by {\small CLOUDY}. 

When the test is initialised, all elements are in their neutral state but as the test proceeds, collisional ionisation transitions the elements to their second ionisation state before finally reaching equilibrium with a dominant third ionisation state (except for Fe and Ne where the fourth and second ionisation states dominate, respectively). It is precisely this evolution that we wish to capture in cosmological simulations.

For our fiducial test, we have used the accurate {\small DVODE} BDF solver. We compare the results from our fast ODE solver (shown as the dashed lines in Figure~\ref{coll_ion_test_0}) with the fiducial model and we find very good agreement for all elements. For our high-resolution cosmological radiative transfer simulations \citep[e.g.][]{Rosdahl2018}, typical time steps are on the order of $10^4{\rm yr}$ which is approximately equal to and often smaller than the steps used in this test\footnote{The time steps in {\small RAMSES-RTZ} are adaptive. Small time steps are taken at the beginning to demonstrate the early evolution of the ionisation fractions and longer time steps are used later in the evolution. As long as the time step is small compared to the equilibrium time scale, the code converges to the correct values.} and hence we expect the results to be equally if not even more accurate for cosmological runs.

A more stringent test of collisional ionisation is to ensure that the code produces the correct equilibrium abundances across a wide range in temperature. We have repeated the same experiment as above; however, in this case, we vary the temperature in the range $10^4{\rm K}-10^8{\rm K}$ and we employ the fast solver. In Figure~\ref{coll_ion_test_1}, we show the ionisation fractions of different oxygen and iron ions as a function of temperature for the test results (solid and dashed lines) versus {\small CLOUDY} (dotted lines). We run the test with our fiducial, reduced chemical model (solid lines) as well as a full chemical model (dashed lines) that uses all ions. At all temperatures we find good agreement between {\small RAMSES-RTZ} and {\small CLOUDY}, confirming the accuracy of our method. The largest difference occurs at temperatures between $10^{6.5}{\rm K}-10^7{\rm K}$ for oxygen ions for the reduced chemical model. The reason for this is because in the reduced chemical model, we artificially truncate the ionisation states of oxygen at O{\small VIII} rather than at O{\small IX}. We will use this method commonly in future simulations in order to reduce the computational load. In this temperature range, small quantities of O{\small VII}, O{\small VIII}, and O{\small IX} are all populated and by artificially truncating the ionisation state at O{\small VIII}, this may slightly overproduce O{\small VII} due to recombinations. However, the difference is clearly insignificant, demonstrating that the method is safe to use in cosmological simulations. We emphasise that this behaviour is to some extent ion-dependent. For example, iron exhibits significantly more overlap of ionisation states at high temperatures. In the bottom panel of Figure~\ref{coll_ion_test_1} we show a similar comparison of ionisation state as a function of temperature for iron. The reduced chemical model (with 10 Fe ions) is applicable only up to $10^6$K. Thus if one wishes to study gas at higher temperatures, one must include more Fe ions. This is particularly important for high temperature gas cooling where high ionisation states of Fe can become important at high metallicities. To demonstrate the accuracy of our model even at these high temperatures, we run an additional experiment including all Fe ionisation states and find very good agreement with {\small CLOUDY}, as expected. There is a small difference between {\small RAMSES-RTZ} and {\small CLOUDY} at $2\times10^5{\rm K}\leq T\leq 2\times10^6{\rm K}$. This can be attributed to the fact that {\small CLOUDY} has updated dielectronic recombination rates for iron that are not included in our code. We also highlight a small difference between {\small RAMSES-RTZ} and {\small CLOUDY} at temperatures slightly higher than $10^4$K. This is due to the fact that the ionisation and recombination rates for H and He that set the electron density are different in {\small RAMSES-RTZ} and {\small CLOUDY}.

\subsection{Photoionisation}
Having demonstrated that our solver is adequate for modelling collisional ionisation, we now move to photoionisation. For the first test, we set the gas density to $10^{-2}{\rm cm^{-3}}$ and the pressure to $P/k_{\rm B}=500\ {\rm cm^{-3}}$K (which corresponds to an equilibrium temperature of $\sim30,000$K). Because collisional ionisation has been artificially turned off for this test, the temperature only plays a role in setting the recombination rates. The test is set up so that the gas is irradiated by a \cite{Haardt2012} UV background assuming the shape and strength at $z=0$. In order to maintain consistency with {\small CLOUDY}, we have output the photoionisation rates for each species directly from {\small CLOUDY} and input them into the simulation. Each cell in the simulation is subject to the same photoionisation rate and the radiation is not propagated. 

In Figure~\ref{phot_ion_test_0} we show the time evolution of the ionisation fractions of the first four ionisation states of O, N, C, Mg, Si, S, Fe, and Ne. In all cases we can see that the equilibrium states reached by the code (solid lines) are in very good agreement with the expectations from {\small CLOUDY} (dotted lines). We once again test that the fast solver (dashed lines) produces the same result as the accurate solver and confirm that this is the case for both the time evolution as well as the final equilibrium abundance.

A more stringent test requires varying the gas density across a range in values. For this reason, we have repeated the experiment (using the fast solver) where we change the gas density between $10^{-3}{\rm cm^{-3}}-10^{3}{\rm cm^{-3}}$. We compare the equilibrium oxygen abundances of the first four oxygen states to the equilibrium values obtained with {\small CLOUDY} in Figure~\ref{phot_ion_test_1}. At all densities the equilibrium results from our code agree with the expectations from {\small CLOUDY}, once again demonstrating the accuracy of our model. 

\begin{figure*}
\centerline{\includegraphics[scale=1,trim={0 0 0 0},clip]{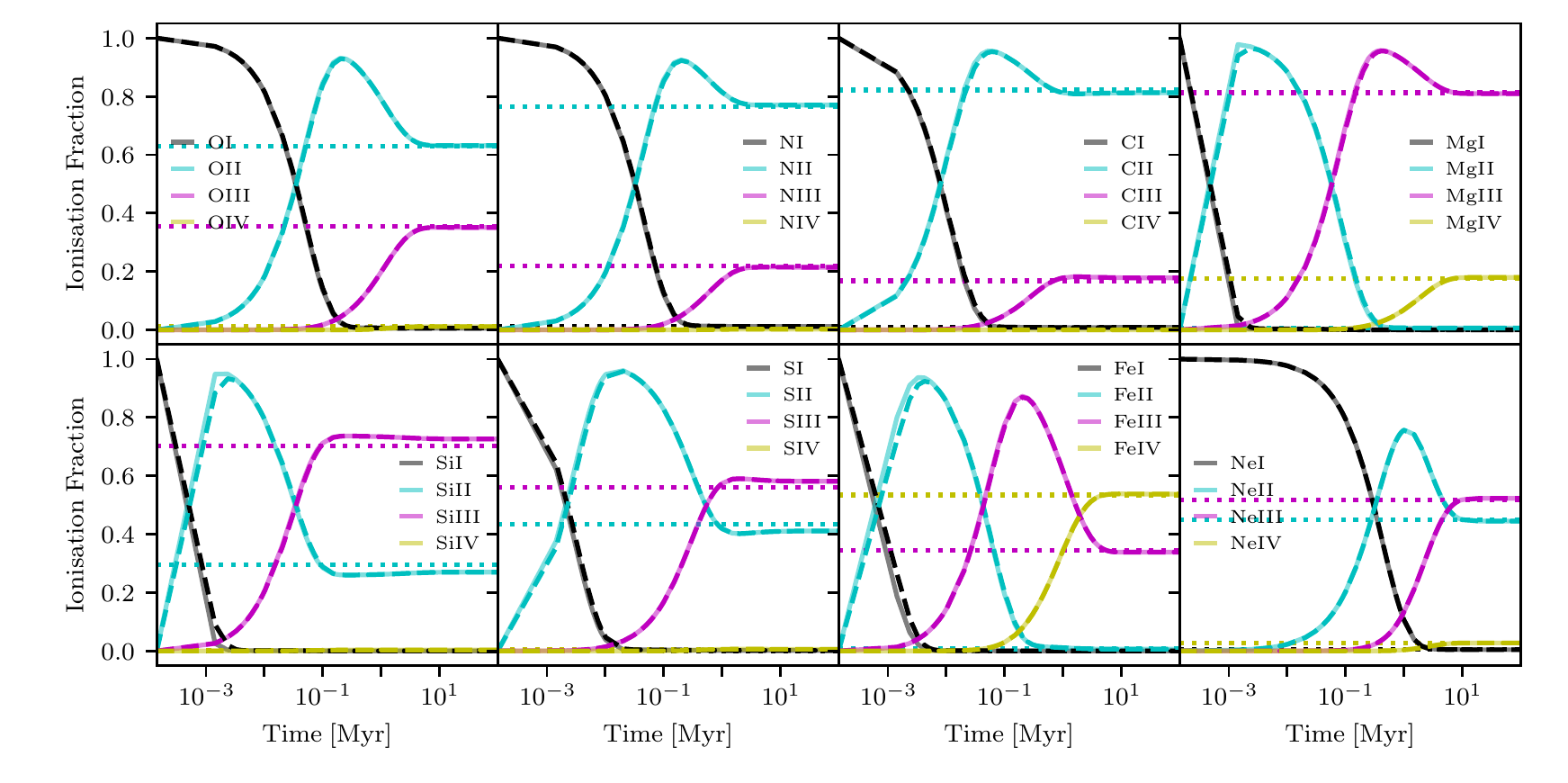}}
\caption{Ionisation fractions as a function of time for the first four ionisation states of O, N, C, Mg, Si, S, Fe, and Ne including both collisional ionisation and photoionisation from a \protect\cite{Haardt2012} UV background at $z=0$. All elements are initialised in their neutral state with the total gas density and pressure set to $10^{-2}{\rm cm^{-3}}$ and $P/k_{\rm B}=500\ {\rm cm^{-3}}$K, respectively. The solid and dashed lines show the results from the test when using the accurate and fast solvers, respectively. The dotted lines represent the equilibrium abundances measured with {\small CLOUDY}. In all cases, the accurate and fast solvers agree with each other and the tests reproduce the equilibrium abundances from {\small CLOUDY} when equilibrium has been reached. }
\label{coll_phot_ion_test_0}
\end{figure*}

\begin{figure*}
\centerline{\includegraphics[scale=1,trim={0 0 0 0},clip]{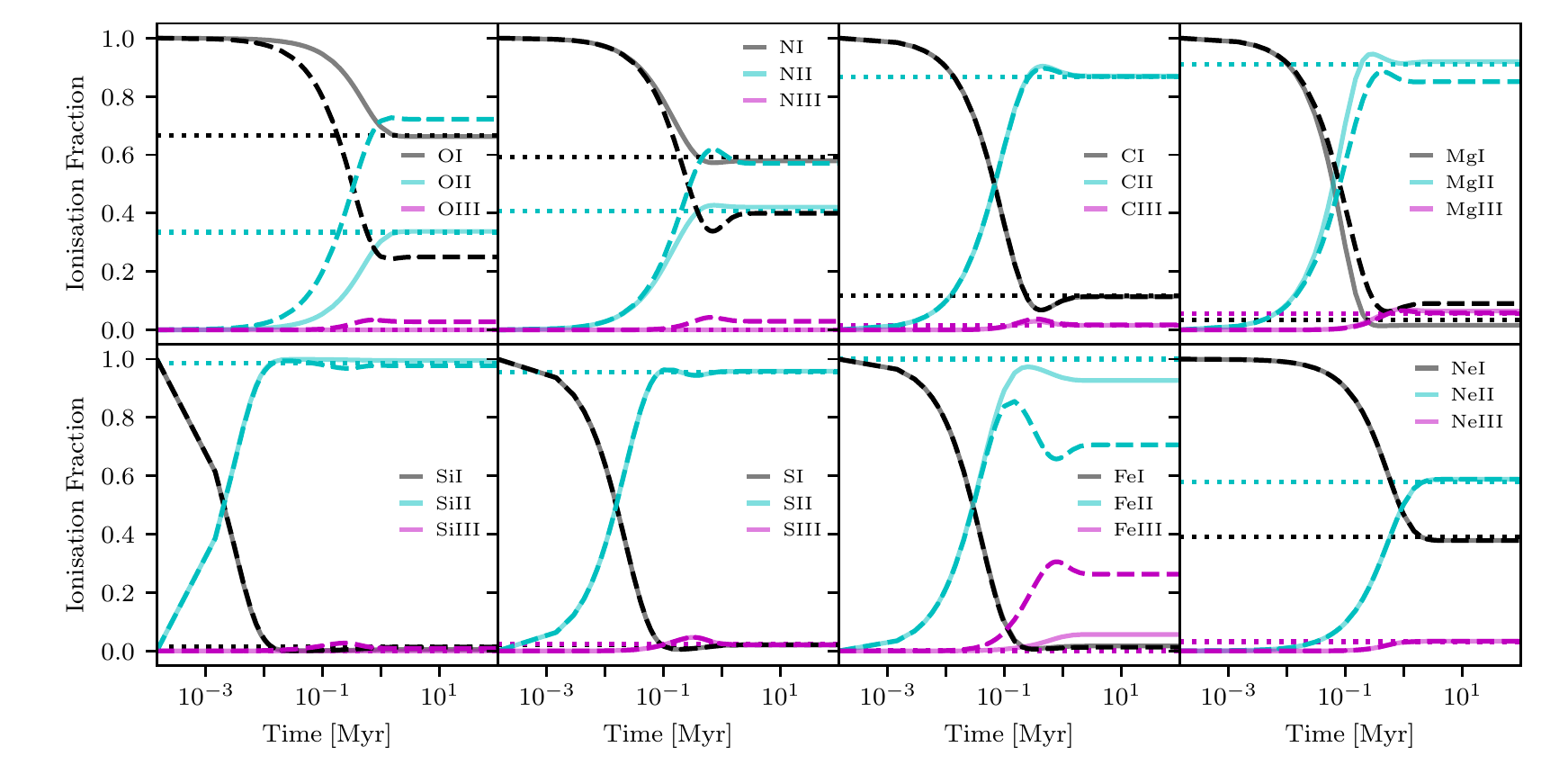}}
\caption{Ionisation fractions as a function of time for the first three ionisation states of O, N, C, Mg, Si, S, Fe, and Ne including both collisional ionisation and photoionisation from a \protect\cite{Haardt2012} UV background at $z=0$. The solid and dashed lines show the results from the tests with and without charge exchange, respectively. All elements are initialised in their neutral state with the gas density and pressure set to $10^{-1}{\rm cm^{-3}}$ and $P/k_{\rm B}=500\ {\rm cm^{-3}}$K, respectively and results are shown for the fast solver. The dotted lines represent equilibrium ionisation fractions measured with {\small CLOUDY}. In all cases, the simulations reproduce {\small CLOUDY} results when equilibrium has been reached and for certain ions (e.g. O, N, and Fe), charge exchange is a very important process. }
\label{ct_phot_ion_test_0}
\end{figure*}

\subsection{Collisional Ionisation \& Photoionisation}
It is clear from the previous two sections that the solver reaches the correct equilibrium ionisation fractions when collisional ionisation and photoionisation are treated independently. We provide one further example demonstrating the accuracy of our solver when both processes are included in the calculation. The setup is identical to the photoionisation test case and the temperature regime is chosen so that both photoionisation and collisional ionisation become important for certain ions. 

In Figure~\ref{coll_phot_ion_test_0} we once again show the ionisation fractions of the first four ionisation states of O, N, C, Mg, Si, S, Fe, and Ne as a function of time. In all cases, the equilibrium values reached by both the accurate solver (solid lines) and fast solver (dashed lines) agree with the expectations from {\small CLOUDY}. Compared to the case where only photoionisation is included, one can see very strong differences in the final Mg ionisation states as well as more subtle differences for O, N, and C. 

As collisional ionisation and photoionisation are the dominant processes that set the ionisation states of individual metals, we use this test to also compare the performance of both the accurate and fast solvers. Although in subsequent sections we will include charge exchange reactions, these break the symmetry of the differential equation matrices and are thus not as viable for our operator split method using the accurate solver. In terms of overall CPU time, the test with the fast solver finished $4.5\times$ faster than the test with the accurate solver. However, a significant portion of the time is spent on the hydrodynamics computation and I/O. Comparing only the chemistry portion of the code, we find that the fast solver is $15\times$ faster than the accurate solver. This speedup is very encouraging for use in cosmological simulations; however, we emphasise that the actual speedup will be dependent on the gas conditions (i.e., temperature and density), local radiation field, simulation time step (which in part dictates the number of sub-steps needed by the accurate solver), and number of ions followed.

\subsection{Charge Transfer}
The final physical process included in our network that impacts the metal ionisation states is charge exchange reactions. These are particularly important at low temperature and high density. To assess the ability of our code to model charge transfer, we set up a cube of gas with a density of $10^{-1}{\rm cm^{-3}}$, a pressure of $P/k_{\rm B}=500\ {\rm cm^{-3}}$K (which corresponds to an equilibrium temperature of $\sim3,500$K), and solar metallicity. In this case, the metallicity is very important due to the dependence of the ionisation states on the number density of each element which appears in the coupling to hydrogen. In order to make an easier comparison with {\small CLOUDY} we have turned off both molecular hydrogen and helium due to the difference in recombination and ionisation rates between the two codes. Like the previous test, we include both collisional ionisation and photoionisation from a $z=0$ \cite{Haardt2012} UV background.

\begin{figure}
\centerline{\includegraphics[scale=1,trim={0 0.0cm 0cm 0cm},clip]{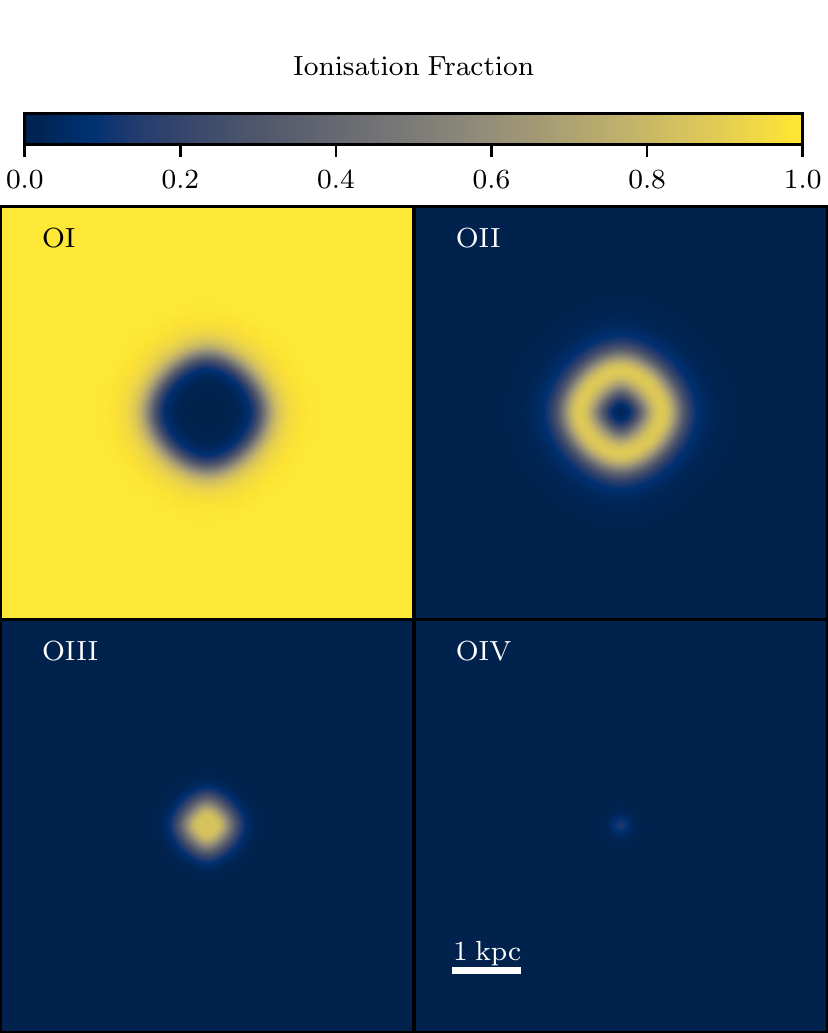}}
\centerline{\includegraphics[scale=1,trim={0 0.0cm 0cm 0cm},clip]{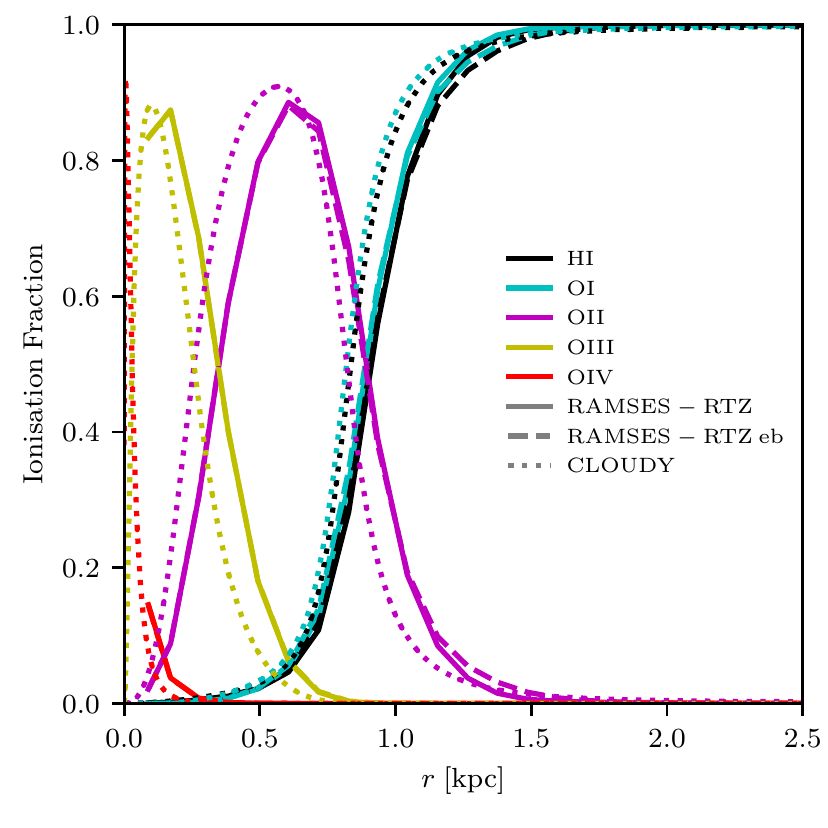}}
\caption{Ionisation fractions of H{\small I} and the first four ionisation states of oxygen as a function of the distance from a 15M$_{\odot}$ zero-age main-sequence population III star in a constant density and temperature medium of $10^{-2}{\rm cm^{-3}}$ and $10^4$K. The simulation is stopped once the Stromgren radius has been reached. The top panel shows thin slices through the centre of the simulated box while the bottom panel shows the radial profile. Solid and dashed lines show the results from {\small RAMSES-RTZ} using the fiducial number of photon energy bins and the enhanced number, respectively. We show the results from {\small CLOUDY} as dotted lines. In all cases, the results qualitatively agree.}
\label{strom1}
\end{figure}

\begin{figure}
\centerline{\includegraphics[scale=1,trim={0 0.7cm 0cm 1.0cm},clip]{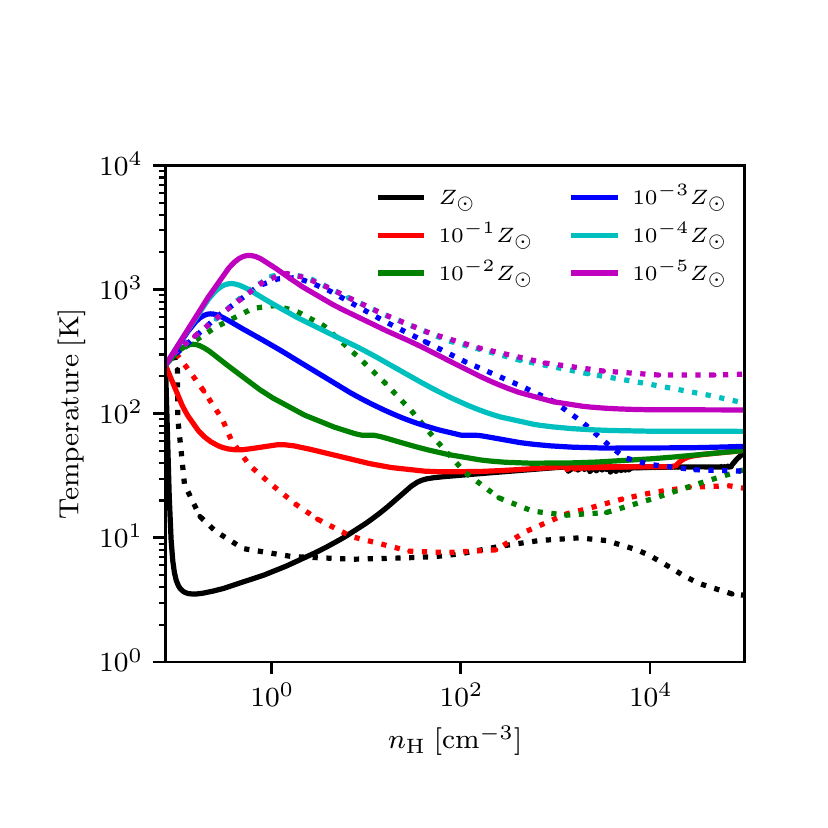}}
\caption{Gas temperature as a function of metallicity for the cloud collapse test at various metallicities. The solid curves show the results from {\small RAMSES-RTZ} while the dotted lines are for a similar experiment from \protect\cite{Omukai2005}. Due to differences in the physics included, the results from \protect\cite{Omukai2005} are not meant to overlap with {\small RAMSES-RTZ}, rather they are shown as a guide for the expected qualitative evolution of the cloud.}
\label{collapse_test}
\end{figure}

\begin{figure*}
\centerline{\includegraphics[scale=1,trim={0 0 0 0},clip]{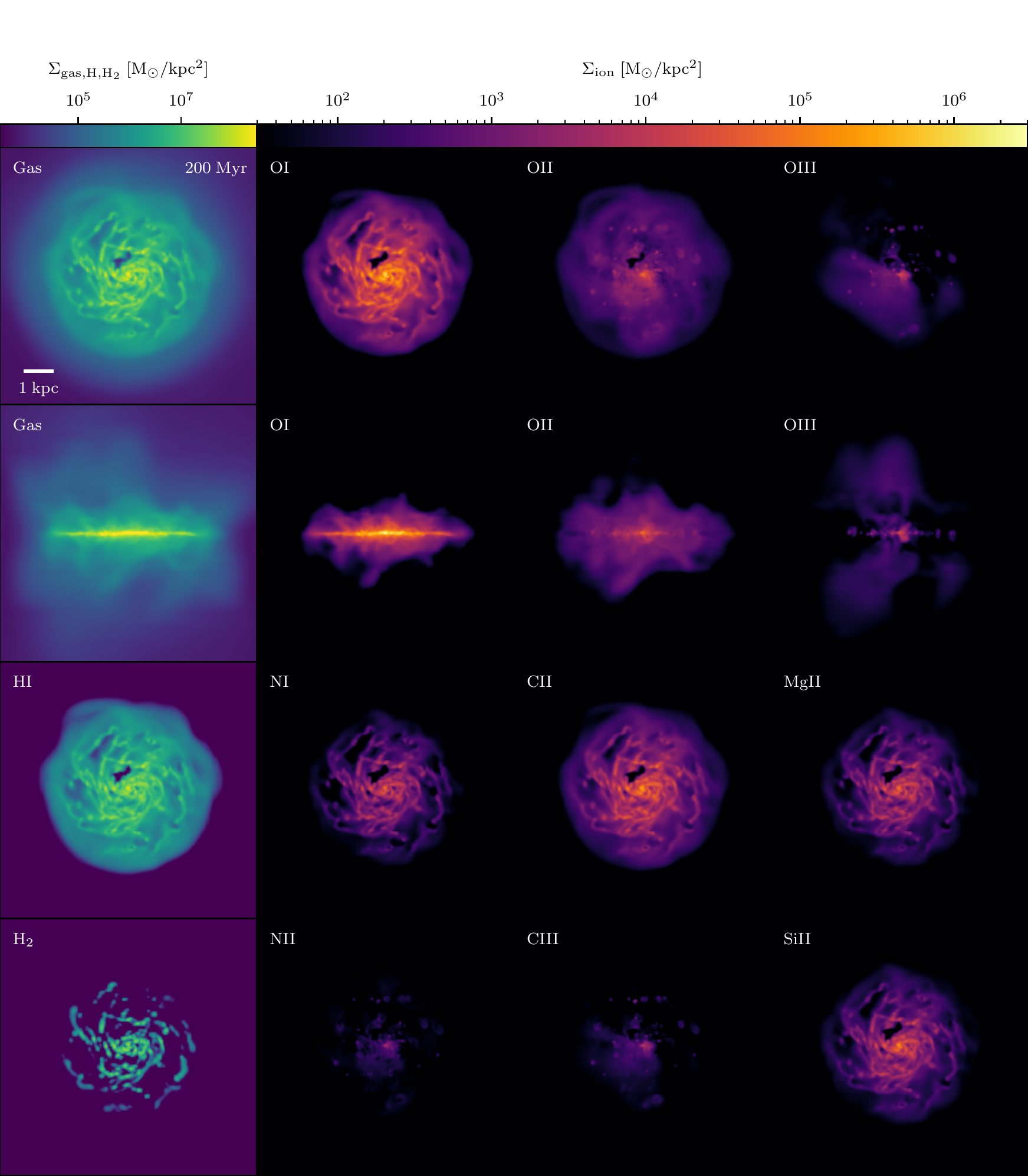}}
\caption{Face-on and edge-on images of the gas surface density and the surface density of various ions as indicated on the images at 200~Myr. The width of each image is 10~kpc. Different ions have different morphologies. For example, O{\small I} is mostly confined to the disc whereas O{\small II} and O{\small III} are present in the outflows.}
\label{G8_O_im}
\end{figure*}

In Figure~\ref{ct_phot_ion_test_0} we show the time evolution of the ionisation fractions of O, N, C, Mg, Si, S, Fe, and Ne as predicted by the fast solver (solid lines) compared with the equilibrium values obtained with {\small CLOUDY} (dotted lines). For all elements the agreement is very good once equilibrium is reached, demonstrating the accuracy of our model. For comparison, we also show the evolution of the ionisation fractions when charge exchange is excluded (dashed lines). As expected \citep{Oppenheimer2013}, excluding charge exchange leads to a significant under-prediction of the O{\small I} abundance. A similar effect, albeit less severe, is seen for N{\small I} and Mg{\small II} while charge exchange is also very important for Fe. Without charge exchange, one would expect that O{\small II} and N{\small II} should dominate when in reality, this is unlikely to be the case at these temperatures and densities. 

We emphasise that obtaining this agreement is non-trivial. The collisional ionisation and recombination equations used in {\small RAMSES-RTZ} differ from those in {\small CLOUDY} for H and He. Similarly, the charge exchange reaction rates have been updated in {\small CLOUDY} for particular reactions (especially those including low ionisation states of oxygen) since \cite{Kingdon1996}. For this particular test, we have output the recombination (using {\small ``save recombination coefficients''}) and ionisation (using {\small ``save ionization rates''}) rates directly from {\small CLOUDY} and ensured that our code is using the exact same values; however, across a range of densities and temperatures we find that our reaction rates differ from those output by {\small CLOUDY} at the level of tens of percent. As described above, for particular oxygen ions, we include the updated charge exchange reaction rate equations from {\small CLOUDY}; however, for all others we rely on \cite{Kingdon1996}. This can lead to small discrepancies for Fe, S and Mg. This is not particularly worrying as the uncertainties for these reaction rates are larger than this difference \citep[e.g.][]{Ferland1998}; however, we highlight this for reproducibility.

\begin{figure*}
\centerline{\includegraphics[scale=1,trim={0 0.6cm 0 0},clip]{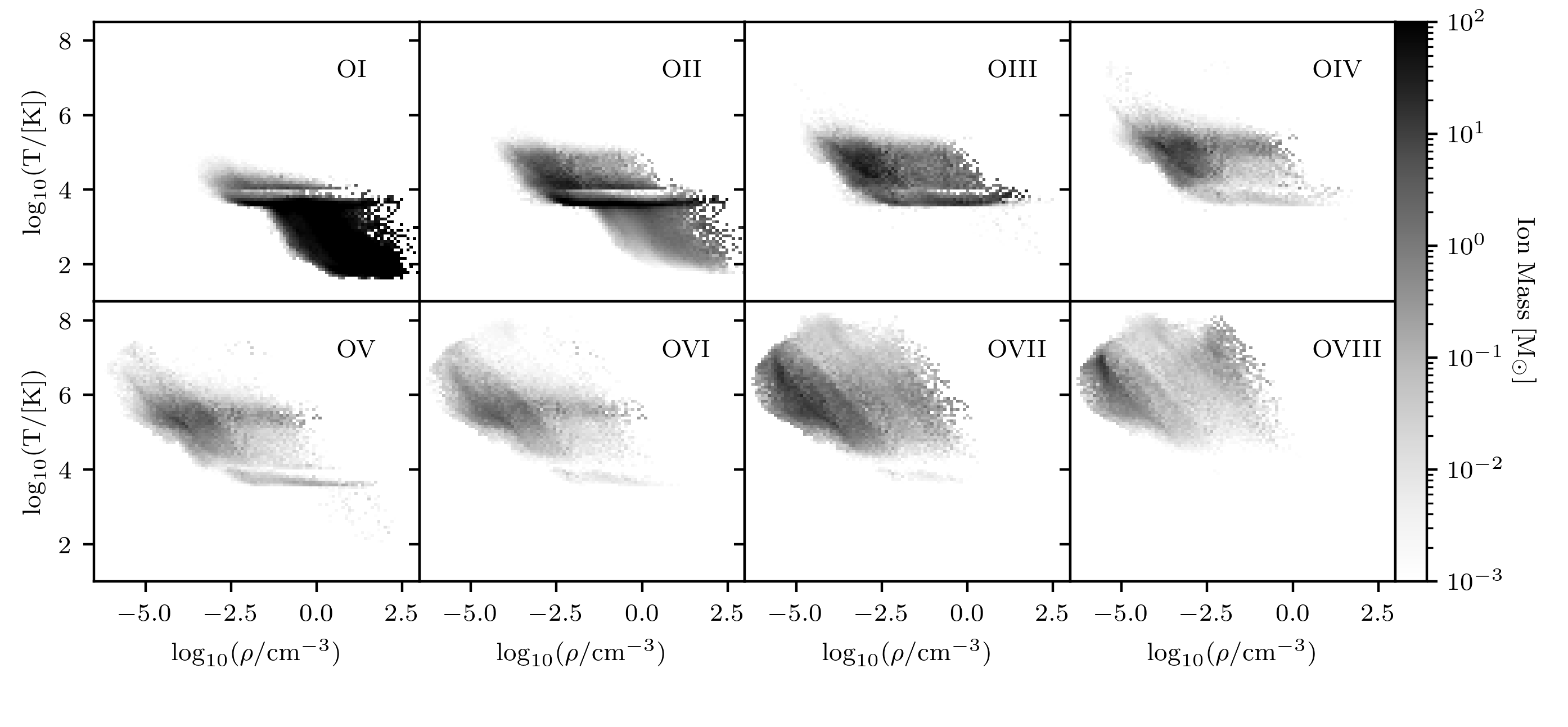}}
\caption{Temperature-density phase-space diagrams of gas in the simulation volume at 200~Myr highlighted by the mass of each oxygen ion that exists at that phase. Higher ionisation states exist at consistently hotter temperatures.}
\label{G8_O_ion}
\end{figure*}

\subsection{On-the-fly Radiative Transfer}
The primary aspect of {\small RAMSES-RTZ} that differentiates it from many other codes in the literature is the ability to couple the metal chemistry with on-the-fly radiative transfer. We test this coupling by setting up a test similar in spirit to Test~1 of \cite{Iliev2006}. We initialise a constant density volume of 6.6$^3$kpc$^3$ with a total gas density of $\rho=10^{-2}{\rm cm^{-3}}$ and temperature of $10^4$K. The gas is set to have Solar metallicity and an initial electron fraction of $10^{-6}{\rm cm^{-3}}$. A source resembling a 15${\rm M_{\odot}}$ zero-age main-sequence population III star \citep{Schaerer2002} is placed at the centre of the box, which emits as a black body with an effective temperature of $10^{4.759}$K and a luminosity of $10^{4.324}L_{\odot}$. The grid in the test is resolved with 64$^3$ cells and we employ our fast solver. For comparison, we set up a similar experiment in {\small CLOUDY} using a spherical geometry with an inner radius of 1pc and an outer radius of 3.3kpc. The outer radius is unimportant as the Stromgren radius is well within the sphere. For ease of comparison, we ignore helium. To be able to make a fair comparison with {\small CLOUDY}, we have turned off cooling in the simulation in both {\small RAMSES-RTZ} and {\small CLOUDY}, neglect molecules, and we run {\small RAMSES-RTZ} in static-mode so that no gas properties other than ionisation state are updated.

Radiation in the simulation is followed in eight energy bins that we commonly employ with {\small RAMSES-RT}. The bins have edges at 0.1eV, 1.0eV, 5.6eV, 11.2eV, 13.6eV, 15.2eV, 24.59eV, 54.42eV, and 500.0eV (see \citealt{Kimm2017}). The first four bins are non-ionising for hydrogen and the latter six cutoffs are set at the Habing band, Lyman Werner Band, H-ionising, H$_2$-ionising, He{\small I}-ionising, and He{\small II}-ionising energies respectively. The first two bins are in the optical and IR and are generally used to follow radiation pressure (which is not included in this experiment).

Two key differences between {\small RAMSES-RTZ} and {\small CLOUDY} are that {\small CLOUDY} can model the SED of the source with many more energy bins and the gas sphere can be broken into much smaller (spatial) zones. The former may impact the ability of {\small RAMSES-RTZ} to model self-shielding and ionisation fronts properly \citep[e.g.][]{Mirocha2012} while the latter limits our ability to model the detailed physics in regions close to the source that are unresolved by the code. The test in this section will illuminate the possible issues with the coupling between the multi-frequency radiative transfer and metal chemistry in {\small RAMSES-RTZ}.

In Figure~\ref{strom1} we show maps of the first four ionisation states of oxygen and profiles of the ionisation fractions of H{\small I}, O{\small I}, O{\small II}, O{\small III}, and O{\small IV} for the test (solid lines) after the Stromgren radius has been reached and compare with the results from {\small CLOUDY} (dotted lines). We find very food qualitative agreement between the two codes for all ions indicating that {\small RAMSES-RTZ} sufficiently captures the physics that governs the evolution of H{\small II} regions. We do however identify certain minor quantitative differences between {\small RAMSES-RTZ} and {\small CLOUDY}. First, the ionisation fraction of O{\small IV} approaches unity very close to the source in {\small CLOUDY}, whereas in {\small RAMSES-RTZ}, it only approaches 20\% in the average profile. This is a severe limitation of the finite spatial resolution employed in the test. The cell widths in the test are $\sim100$pc while the smallest zones in {\small CLOUDY} are $\sim10^{-2}$pc. The most aggressive modern cosmological zoom-in simulations that follow the formation of individual galaxies have spatial resolutions that are still orders of magnitude larger than the smallest zone sizes in {\small CLOUDY}. This must be kept in mind when interpreting high ionisation lines emitted very close to sources. We also note other small discrepancies between the other ion profiles as a function of radius. In general, the ion fractions peak slightly earlier for {\small CLOUDY} compared to {\small RAMSES-RTZ}. Resolution can certainly be an issue here as well but we have run lower resolution simulations with {\small RAMSES-RTZ} (i.e. 32$^3$ cells) and found only small differences in the results. Another possibility is that the differences in SED resolution are impacting self-shielding and the photoionization rates. We test this by rerunning the test with additional photon energy bins with the following boundaries: 0.1eV, 1.0eV, 5.6eV, 11.2eV, 13.6eV, 15.2eV, 17.5eV, 20.0eV, 24.59eV, 30.0eV, 35.12eV, 54.42eV, 54.94eV, 138.1eV, and 500.0eV. Many of these new energy bins were selected so that the ionisation energies of various oxygen ions no longer fell in the middle of a photon energy bin. We show the results of this test as the dashed lines in Figure~\ref{strom1}. Indeed we find that the results with finer energy sampling for the SED agree almost perfectly with the original test using the fiducial photon energy bins, indicating that the self-shielding and photoionization rates are well captured even with our fiducial coarse sampling. Thus it is likely that other differences between {\small RAMSES-RTZ} and {\small CLOUDY} are responsible for the different ionisation profile shapes.

We have investigated this further and found mild differences between the total photoionisation cross sections between {\small RAMSES-RTZ} and {\small CLOUDY}. As described above, we attempt to maintain as close of an agreement as reasonable with the atomic data used in {\small CLOUDY}; however, certain differences persist that prevent a perfect match of our test results. We thus attribute the differences not associated with spatial resolution issues to atomic data discrepancies rather than SED sampling. Additionally, we note that the quantitative differences between {\small RAMSES-RTZ} and {\small CLOUDY} are in line with those found between {\small CLOUDY} and other PDR codes \citep{Rollig2007}.

One of the key features of this experiment is that the location of the O{\small I} ionisation front almost perfectly matches that of H{\small I}. Without the inclusion of charge exchange, the O{\small I} ionisation front is expected to reside further from the source than the H{\small I} ionisation front. However, rapid charge exchanges force them into equilibrium. The importance of charge exchange for low ionisation states of oxygen was highlighted earlier and this additional test reflects the ability of {\small RAMSES-RTZ} to capture this well known effect \citep[e.g.][]{Abel2005} with on-the-fly RT.

\subsection{One Zone Cloud Collapse}
As we have introduced non-equilibrium metal line cooling into {\small RAMSES-RTZ}, we provide one final benchmark where we model the collapse of a prestellar cloud at low temperatures and densities. The test follows the methodology in \cite{Omukai2005}. A single cell is initialised with a density of 0.1cm$^{-3}$ and a temperature of $300$K in an entirely neutral configuration at various metallicities in the range $10^{-5}Z_{\odot}-Z_{\odot}$. The density is evolved such that
\begin{equation}
    \frac{d\rho}{dt}=\frac{\rho\sqrt{1-f}}{\sqrt{\frac{3\pi}{32G\rho_0}}},
\end{equation}
where $\rho_0$ is the initial gas density and $f=0.744$, appropriate for a gas with $\gamma=5/3$ (see Equation 9 of \citealt{Omukai2005}). The temperature is updated by solving for the evolving internal energy of the system including metal line cooling, using all of the available lines in {\small RAMSES-RTZ} as listed in Section~\ref{sec:cooling}, as well as primordial cooling from H, He, and H$_2$ (see Equations 1, 2, \& 3 of \citealt{Omukai2005}). The cell is evolved until it reaches a density of $10^5{\rm cm^{-3}}$.

In Figure~\ref{collapse_test} we show the evolution of the gas temperature as a function of density for the test. For the low-metallicity clouds, the temperature increases initially due to compressional heating. However, as the density increases, so does the strength of the cooling; hence the turnover in temperature. As the metallicity increases, the peak temperature occurs at a lower value and at lower densities. For the most metal enriched clouds (i.e. $10^{-1}Z_{\odot}$ and $Z_{\odot}$), the initial cooling dominates over the compression heating and thus the temperature drops rapidly from the beginning of the test.

The dotted lines in Figure~\ref{collapse_test} show the results \cite{Omukai2005}. Because of differences in cooling and chemistry\footnote{For example, \cite{Omukai2005} includes cooling from molecules such as CO, OH, and H$_2$O, heating and cooling from dust grains, a chemical network of 50 species and nearly 500 reactions (most of which are not included in {\small RAMSES-RTZ}), a more sophisticated model of H$_2$ cooling, and a model for optical depth effects on the cooling rate.}, the lines are not expected to overlap with ours, rather we show them to demonstrate the expected qualitative evolution of the system. Indeed \cite{Omukai2005} also find that for the highest metallicity collapse tests, cooling always dominates over compressional heating (see also \citealt{Grassi2014}) and the peak temperatures reached in the lower metallicity runs agree to within a factor of two.

\section{Example Simulation}
\label{otf}
Thus far, our benchmark tests have only demonstrated that the solver produces the correct time evolution and equilibrium ionisation fractions in a static configuration. In this section, we demonstrate an example of how the code is applied to dynamic environments where non-equilibrium chemistry and cooling are computed at a variety of temperatures and densities and radiation is advected simultaneously.

We simulate the evolution of an isolated dwarf galaxy \citep[G8,][]{Rosdahl2015b} that has a circular velocity of $\sim30{\rm km\ s^{-1}}$ using our new non-equilibrium chemistry and cooling scheme. Details of the initial conditions can be found in \cite{Rosdahl2015b,Kimm2018}. The simulation is initialised to have a mass-weighted disc metallicity of $\sim0.13Z_{\odot}$ with abundances following \cite{Grevesse2010} and a metal-free halo. The initial metallicity exhibits a gradient across the disc such that $Z=0.1Z_{\odot}10^{0.5-r/5.0{\rm kpc}}$. Furthermore, all ions are initialised in their ground state in the disc and in their most ionised state that we track in the halo. In addition to H and He ionisation states as well as H$_2$ \citep{Katz2017}, for this simulation we follow eight ionisation states of O, seven of N, six of C, and six of Si, Mg, Fe, S, and Ne. We employ our ``fast'' solver.

Stars are formed using a thermo-turbulent star formation criteria \citep[e.g.][]{Federrath2012,Kimm2017}. The simulation includes relevant stellar feedback processes such as supernova (SNII and SNIa) explosions which are accounted for using the mechanical feedback prescription of \cite{Kimm2015} and stellar winds \citep{Agertz2021}. We follow the chemical enrichment from individual SNe explosions based on mass and metallicity as well as stellar winds using yields from \cite{Portinari1998} for SNII, \cite{Seitenzahl2013} for SNIa, and \cite{Pignatari2016} for AGB winds. Stars also inject radiation into their host cells based on their mass and metallicity following a {\small BPASS} SED \citep{Stanway2016,Eldridge2008} and the radiation is tracked in eight radiation bins from the IR to the UV \citep[see][]{Kimm2018}. In addition to the stellar radiation, the box is permeated by a $z=0$ UV background \citep{Haardt2012} and we apply a simple self-shielding prescription for the background where we exponentially decrease the intensity of the background at densities $\geq10^{-2}{\rm cm^{-3}}$. The radiation impacts the gas via photoionisation, radiation pressure, and photo-heating. Cross sections for the individual ions are computed on-the-fly and updated every 10 coarse time steps in the simulation to be consistent with the luminosity weighted average of all of the radiating star particles. For computational efficiency, we use the reduced speed of light approximation setting the value of $c_{\rm reduced}$ in the simulation to $c/100$. 

Gas cells are refined in the simulation up to a maximum spatial resolution of 18~pc\footnote{At this spatial resolution, Stromgren spheres around individual star particles are not always completely resolved. When this occurs, temperatures for partially ionised cells are not fully representative of and often lower than what the gas temperature would be in a resolved Stromgren sphere. We account for this by identifying all gas cells that host star particles where the combined luminosity of the hosted star particles leads to a Stromgren radius that is less than half the cell width (indicating an unresolved Stromgren sphere). For these cells we use a temperature floor of $10,000$K when computing line emission that originates in ionised regions (e.g., recombination emission for H$\alpha$ and H$\beta$, emission lines from O{\small II}, and O{\small III}, etc.). This has the effect of decreasing the Balmer emission due to the reduced recombination emissivity with increasing temperature while generally increasing nebular metal line emission due to the increasing effective collision strength with temperature.} if their dark matter or gas mass reaches 8 times its initial value, or to resolve the local Jeans length by at least four cells. Because the simulation does not include gas inflows, we only follow the evolution for the first 200~Myr to avoid unrealistic gas metallicities that could result modelling metal enrichment processes without dilution.

In Figure~\ref{G8_O_im}, we show face-on (first row) and edge-on (second row) images of the gas surface density and surface density of three oxygen ionisation states after 200~Myr of evolution. O{\small I} is primarily confined to the disc in the dense neutral regions. This is not surprising considering the ionisation energy of O{\small I} is nearly identical to that of H{\small I}. Probing a more highly ionised state, O{\small II} has a significantly different morphology than O{\small I}. The highest surface  densities of O{\small II} also occur in the disc but are most prominent in star-forming regions where ionising photons can excite oxygen into this state. From the edge-on view, it is clear that O{\small II} is present well above the disc plane, both in outflows as well as gas that is recombining and raining back onto the disk. From the same view we see that O{\small III} exists even further from the plane of the disc as expected. As there are few ionising photons with $E>35$eV needed to ionise O{\small II}, O{\small III} primarily forms via collisional ionisation which occurs in regions heated by SN feedback.

This can be better observed in Figure~\ref{G8_O_ion} where we show temperature-density phase-space diagrams illuminated by the mass of each oxygen ion present. We can confirm from Figure~\ref{G8_O_ion} that O{\small I} primarily populates dense, cold, neutral gas and as we move to each successively more ionised state, the regions that are bright are at progressively higher temperatures and lower densities. The vast majority of O{\small III} is present in gas at temperatures in the range $10^4$K$-10^5$K. Depending on the element, the phase-space diagrams will be illuminated in different regions. We show images of various ions other than oxygen in Figure~\ref{G8_O_im} where one can see the different morphologies of each ionisation state.

One of the key effects that can be captured by simulations that follow non-equilibrium metal ionisation states is where a highly ionised ion, such as O{\small III}, can exist at much lower temperatures than one would expect from, for example, assuming CIE. This can occur when the recombination timescale is longer than the cooling time. This results in an ionisation lag (i.e., the gas is more ionised than expected from equilibrium) as discussed by, e.g., \cite{Kafatos1973,Gnat2007,Oppenheimer2013,Vasiliev2013}, and demonstrates the importance of non-equilibrium effects. Such an effect can be observed in Figure~\ref{G8_O_ion} where we see high oxygen ionisations states present in low-density ($\rho=10^{-1}{\rm cm^{-3}}$), low-temperature ($T\lesssim10^4$K) gas.

One of the primary advantages of the code is that observable quantities such as emission lines and absorption lines/column density distributions are trivial to generate. The latter is simply a line-of-sight integral, complicated only by the velocity structure of the gas. Such studies are well represented in the literature \citep[e.g][]{Oppenheimer2018} so we refrain from discussing them here. For the former, there are numerous methods that have been applied in the literature. One of the most simple methods is to associate emission lines with individual star particles and use the ``out-of-the-box'' values computed for an idealised set of photoionisation models that are often shipped with stellar SEDs \citep[e.g.][]{Ceverino2019,Kannan2021}. Such methods are necessary if the local ISM is unresolved; however, other simulations which suffer similar resolution issues have attempted photoionisation modelling with parameters tuned to those in the simulation \citep[e.g.][]{Wilkins2020}. Unfortunately, the lack of ISM resolution and equilibrium assumption prohibits their use as ISM diagnostics. Higher resolution simulations often apply individual photoionisation models on a ``cell-by-cell'' basis \citep[e.g][]{Katz2020b,Lupi2020}; however, once again, the equilibrium assumption can cause systematic biases \citep{Lupi2020}. Since our simulations follow the non-equilibrium metal ionisation states, electron density, temperature, and radiation field, it now becomes trivial to solve for the level populations and to compute intrinsic emission line luminosities (which is essentially the same calculation that we already do for the non-equilibrium cooling, as discussed in Section~\ref{methods}). To retrieve emission lines that are not part of our cooling model, we use {\small PyNeb} \citep{pyneb2015} to solve for the level populations and emissivities.

Simulators often asses the success of their model by comparing to one-point statistics, for example, the stellar mass function or the stellar mass-halo mass relation derived from abundance matching. Neither of these metrics are actually observed quantities and thus we demonstrate a few examples of how to better compare simulations with observations using emission lines that are by-products of our new model.

\subsection{Star Formation Rate Indicators}
Emission lines, in particular H$\alpha$ and [O{\small II}], are one of the primary methods used to measure SFRs in galaxies \citep[e.g.][]{Kennicutt1998,Kewley2004}. Calibrating these relations from theory is highly non-trivial because there are dependencies on the stellar IMF, stellar metallicity, SED, escape fraction, and gas metallicity among others. For example, the H$\alpha$-SFR relation derived using a {\small BPASS} SED \citep{Eldridge2008} is different from what is obtained from that using {\small STARBURST99} \citep{Leitherer1999}. Similarly, the calibration shifts when varying between a Salpeter, Kroupa, and Chabrier IMF. Furthermore, \cite{Tacchella2021} recently showed that additional corrections to standard scaling relations are needed for the absorption of Lyman continuum photons by dust and helium. Since simulations, in principle, can capture many (but not all) of these effects, we argue that they can be used to calibrate SFR-line luminosity relations. Likewise, when large observational samples of galaxies are available, comparing the H$\alpha$ and [O{\small II}] directly from simulations with other tracers of stellar mass or metallicity may be a more robust test of galaxy formation than using derived quantities from observations based on more simplistic assumptions. 

\begin{figure}
\centerline{\includegraphics[scale=1,trim={0 0 0 1.2cm},clip]{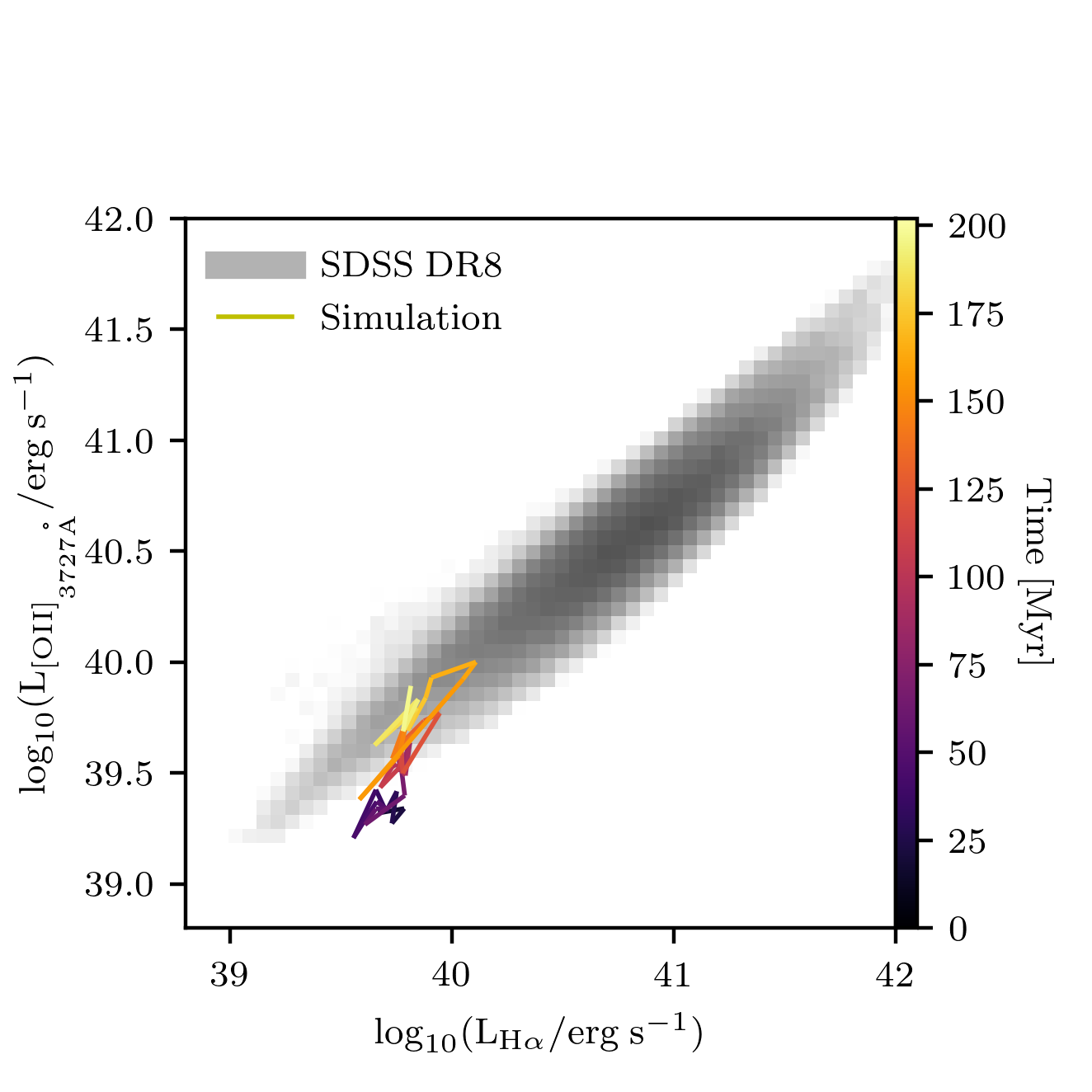}}
\caption{H$\alpha$ versus [O{\small II}] as a function of time (coloured line) for our simulation compared to star-forming galaxies from SDSS (grey histogram). The colour of the line represents the time in the simulation as indicated by the colour bar.}
\label{sfr_rel}
\end{figure}

In Figure~\ref{sfr_rel}, we compare the H$\alpha$ luminosity of the simulated galaxy as a function of time with the [O{\small II}] luminosity. As the galaxy evolves, the luminosities scatter by a factor of few in each line yet their ratio is consistent with the distribution of low-redshift SDSS galaxies \citep{Thomas2013} towards the end of the simulation, when the isolated disc has settled. This indicates that our simulation includes a reasonable (but not necessarily correct) physical model of the ISM since these emission lines are computed directly from the abundances of each ion.

\begin{figure}
\centerline{\includegraphics[scale=1,trim={0 0 0 1.2cm},clip]{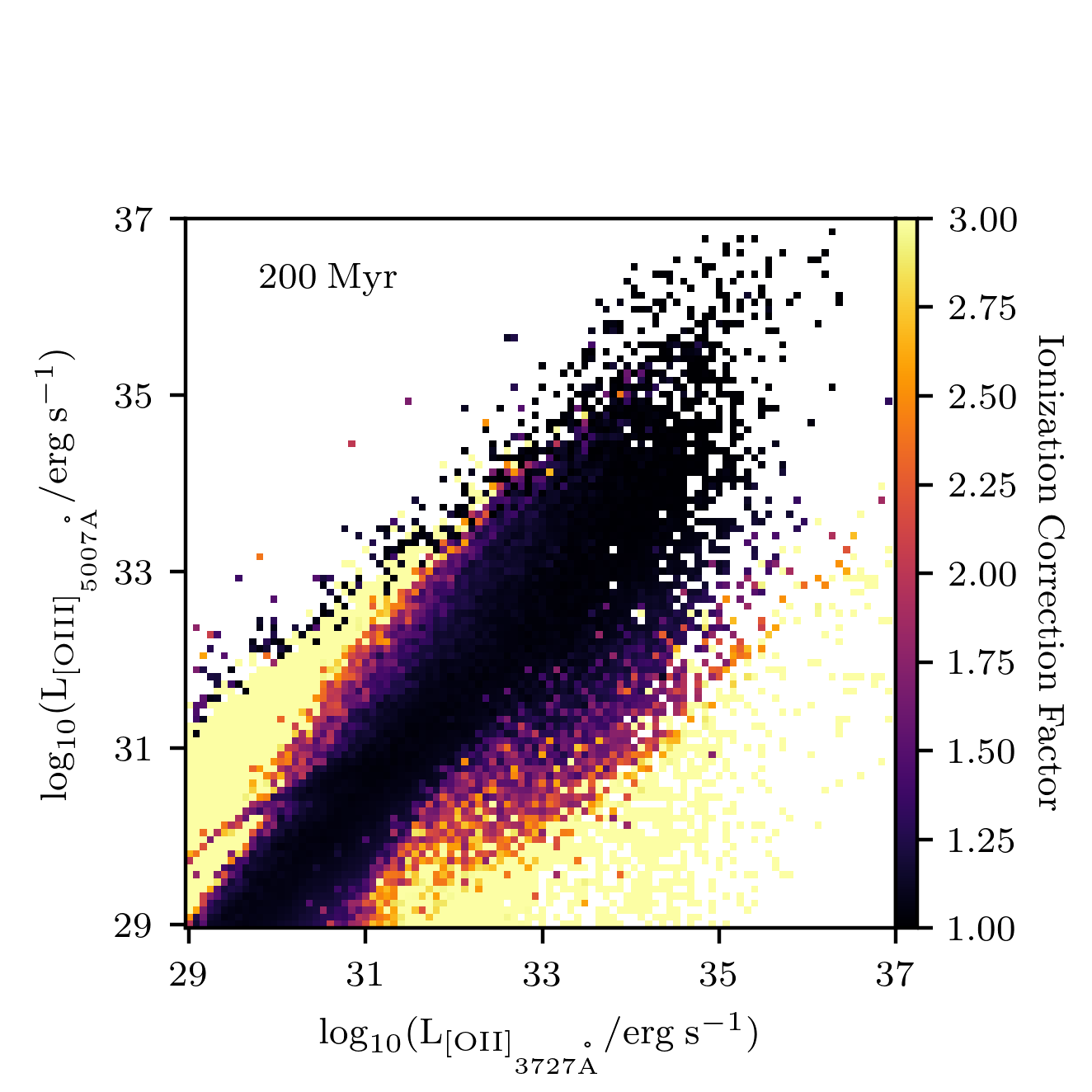}}
\caption{Histogram of [OII] 3727\AA\ emission versus [OIII] 5007\AA\ emission coloured by the mean ionisation correction factor.}
\label{icf}
\end{figure}

\subsection{Metallicity Indicators}
Understanding the stellar mass-metallicity relation and its scatter is one of the major open questions in galaxy formation. Various simulations have taken different approaches in terms of how they compare gas-phase metallicity with observed values. For example, \cite{Torrey2019} measure an average metallicity for a galaxy by summing the total metal content (and scaling appropriately by the oxygen fraction when needed), whereas \cite{Ma2016} describe three different approaches by selecting gas in a fixed temperature range or with a specific ionisation fraction. There are various observational diagnostics of metallicity (see e.g. \citealt{Izotov2006,Curti2017}), many of which rely on specific emission line luminosities but none of which are comparable with the majority of the methods used in simulations. Comparing directly with the emission line diagnostic once again provides a less model-dependent way of testing a galaxy formation model.

Observational metallicity diagnostics based on emission line luminosities are problematic for simulations that do not follow non-equilibrium metal ionisation states for at least two reasons. First, the emission lines only probe specific gas phases (see Figure~\ref{G8_O_ion}). Thus, summing the total metallicity of a galaxy \citep[e.g.][]{Torrey2019} in a simulation may give a metallicity estimate that is based on different gas from what observers actually probe. Selecting gas based on temperature and ionisation thresholds \citep[e.g.][]{Ma2016} may reduce the bias to some extent but in reality, the metallicity obtained from observations is really a complex average of the emission line luminosity-weighted values of the gas probed by each emission line in the diagnostic. Second, depending on the emission lines used, the metallicity probed by different emission lines only captures the metal abundance in the ionisation state needed to create the line. Emission lines from O{\small II} and O{\small III} are commonly used as metallicity diagnostics which means that the oxygen abundance measured by observations either needs an ionisation correction factor \citep[e.g.][]{Peimbert1969}, which can only be estimated by detailed modelling \citep[e.g.][]{DI2014} but comes with large uncertainties, or the abundance that is listed is only for the relevant ionisation states, which is once again, very different from what most simulations measure. 

Following the non-equilibrium metal ionisation states can solve both of these issues. In Figure~\ref{G8_O_ion}, we can see the regions of temperature-density phase-space that are bright in [O{\small III}] and [O{\small II}] and weight the metallicity in the simulation accordingly (the same can be done for H$\alpha$ and H$\beta$). Furthermore, we can better understand the impact of ionisation correction factors directly from the simulation. 

In Figure~\ref{icf}, we show an example of how our new method can be used to reduce the discrepancies between observed and simulated quantities by plotting a 2D histogram of [O{\small II}] versus [O{\small III}] where the pixel brightness represents the mean ionisation correction factor (ICF, defined as the inverse of the sum of the O{\small II} and O{\small III} fractions in the cells). The novelty here is that the ICF is computed directly from the simulation rather than relying on idealised photoionisation models. When both [O{\small II}] and [O{\small III}] are especially bright, the ionisation correction factor approaches unity, which is expected as this gas should be dominated by  O{\small II} and O{\small III}; however, large deviations are seen when only one line dominates the luminosity. Once again, with a large sample of simulated galaxies, the ICF can be studied in a variety of environments and used to better translate between observed emission lines and unobservable quantities such as galaxy metallicity. 

\subsection{Distinguishing Ionisation Mechanisms}
In our final example, we show how our new method can be used to better understand ionisation sources within galaxies. The BPT diagram \citep{Baldwin1981} has traditionally been used to classify sources where the nebular emission is dominated by star formation or an accreting black hole \citep[e.g.][]{Kewley2001,Kauffmann2003}. Trend lines for this emission line diagnostic are highly sensitive to galaxy characteristics such as ionisation parameter, metallicity, gas density, and stellar SED, which are not known {\it a priori}. Within the same galaxy, properties of H{\small II} regions may be different and likewise, there is strong observational evidence that ISM properties are changing with redshift \citep[e.g.][]{Kewley2013}. Cosmological simulations can capture many of these effects and hence provide a better understanding of how the BPT diagram and other strong-line diagnostics can be used to constrain galaxy properties. 

\begin{figure}
\centerline{\includegraphics[scale=1,trim={0 0 0 1.2cm},clip]{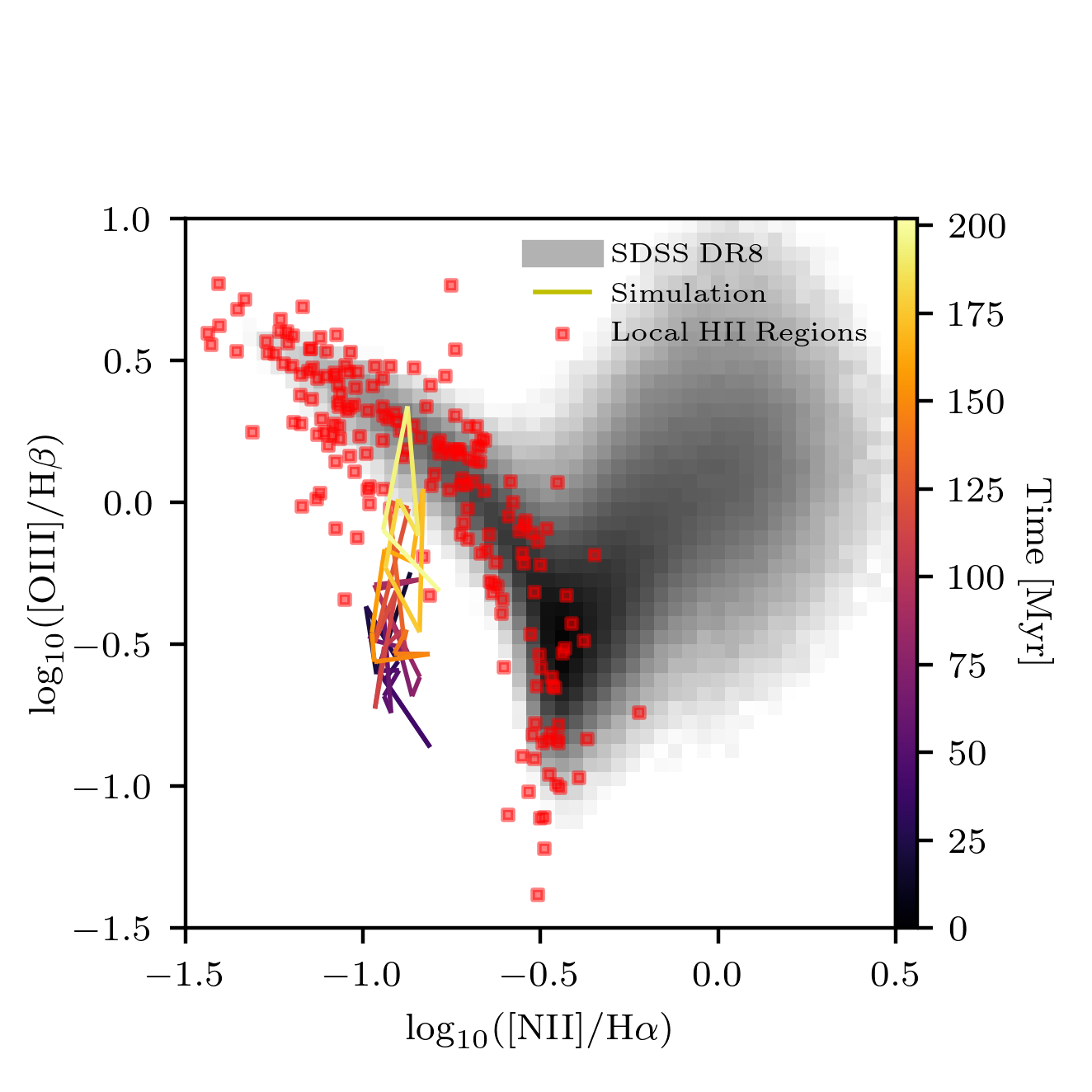}}
\caption{BPT diagram of low-redshift SDSS galaxies (grey histogram) and local HII regions \protect\citep{vanZee1998} versus our simulated, isolated galaxy. The colour of the line represents the time in the simulation as indicated by the colour bar.}
\label{bpt}
\end{figure}

In Figure~\ref{bpt} we demonstrate our ability to predict the evolution of a galaxy on the BPT diagram without relying on external photoionisation models. The galaxy exhibits a significant amount of scatter during the course of its evolution, particularly in the [OIII]/H$\beta$ ratio. Initially, the galaxy falls far away from both the SDSS galaxies \citep{Thomas2013} as well as local HII regions \citep{vanZee1998}, but as the disc settles, it moves into a region more consistent with observations, similar to what was seen in Figure~\ref{sfr_rel}. We emphasise here that the evolution of the galaxy is sensitive to numerous individual H{\small II} regions that combine to produce emission line-ratios as seen in the figure.

We highlight that the physics of this diagram is significantly harder to predict than emission line-ratios that use lines of similar ionisation potentials. O{\small I} and H{\small I} have very similar ionisation potentials and hence we expect to find H$\alpha$ and [OII] emission from approximately the same regions in the galaxy. In contrast, O{\small III} will not extend as far away from the star as the entire H{\small II} region. Thus the spatial resolution in the simulation is key for making robust predictions for emission line diagnostics that use line ratios where ions have large differences in ionisation potentials. This issue is further discussed in Section~\ref{cavs} as the results are very sensitive to the correction for unresolved HII regions.

\subsection{Comparison with Equilibrium Models}
\label{modcomp}
Simulations that do not include non-equilibrium metal chemistry often assume some form of equilibrium in order to calculate emission line luminosities. Here, we compare the results from our non-equilibrium calculation (without corrections for unresolved HII regions) using {\small RAMSES-RTZ} with various estimates for bright oxygen emission lines. We consider the following models.
\begin{itemize}
    \item {\bf Model A}: Collisional ionisation equilibrium for H, He, and metals. These are the values one might obtain when running a simulation without coupled radiative transfer and no UV background.
    \item {\bf Model B}: Collisional and photoionization equilibrium for H, He, and metals. We assume a $z=0$ \cite{Haardt2012} UV background with a self-shielding approximation. These are the values one might obtain when running a simulation without coupled radiative transfer but with a UV background.
    \item {\bf Model C}: The same as Model B but the photoionization rate is directly calculated from the simulation including the UV background and the photon number densities of each cell. These are the values one might obtain when running a simulation with on-the-fly radiative transfer and a homogeneous UV background, as is common in RT-zoom simulations. This is most similar how we have post-processed simulations in the past with {\small CLOUDY} \citep[e.g.][]{Katz2019,Katz2020,Katz2021} assuming a constant temperature model. In other words, the non-equilibrium radiation field is used but equilibrium abundances are calculated for all species as fixing the electron and hydrogen ion densities is not recommended in {\small CLOUDY}.
    \item {\bf Model D}: Collisional and photoionization equilibrium for metals and non-equilibrium values for primordial species and electrons. We use the photoionization rates directly calculated from the simulation including the UV background and the photon number densities of each cell. These are the values one might obtain when running a simulation with coupled radiative transfer and a UV background, but in contrast to Model~C, we fix the primordial species abundances to their non-equilibrium values. This method is likely to be the most accurate since it uses the most information directly from the simulation. 
\end{itemize}
Note that in all models, we use the temperature calculated by the simulation to compute the ionisation and recombination rates (as well as charge transfer coefficients). In reality the temperature one might measure in a simulation that does not follow non-equilibrium metal line cooling is different from the values in the simulation. Thus the luminosity differences between Models A, B, C, and D and the simulation should be considered as minimum differences.

\begin{table}
    \centering
    \caption{Estimated [O{\small II}]$_{3726\angstrom}$, [O{\small II}]$_{3728\angstrom}$, and [O{\small III}]$_{5007\angstrom}$ luminosities of the galaxy disc at 200~Myr for the four models listed in Section~\ref{modcomp}. All luminosities are provided with units of $\log_{10}(L/{\rm erg\ s^{-1}})$.}
    \begin{tabular}{lccc}
        Model & [O{\small II}]$_{3726\angstrom}$ & [O{\small II}]$_{3728\angstrom}$ & [O{\small III}]$_{5007\angstrom}$ \\
        \hline
        Fiducial & 38.28 & 38.43 & 38.96 \\
        \hline
        Model A & 37.95 & 38.12 & 37.45 \\
        Model B & 37.94 & 38.11 & 37.45 \\
        Model C & 38.87 & 38.97 & 38.81 \\
        Model D & 38.21 & 38.35 & 38.74 \\
        \hline
    \end{tabular}
    \label{tab:model_comp}
\end{table}

In Table~\ref{tab:model_comp} we list the [O{\small II}]$_{3726\angstrom}$, [O{\small II}]$_{3728\angstrom}$, and [O{\small II}]$_{5007\angstrom}$ luminosities of the simulation at 200~Myr for each of the models. Models A and B, which assume equilibrium abundances for primordial species but neglect the local radiation field, under predict all three emission line luminosities. This is because the vast majority of the emission in our simulation originates in the Stromgren spheres around star particles and by neglecting this radiation, these regions are not well modelled which leads to large inaccuracies in the line luminosity predictions.

In contrast, when the local radiation is included, line luminosity predictions become significantly more accurate. The results for Model~C are in better agreement with the expectation from the simulation compared to Models~A and B. However, the [O{\small II}] emission line luminosities are over predicted by a factor of $\sim4$ while we find that [O{\small III}] is slightly under predicted. When switching to the non-equilibrium values for primordial species along with the local radiation field as in Model~D, the [O{\small II}] emission line luminosities significantly improve while the [O{\small III}] luminosity becomes marginally worse. It is likely a coincidence that Model~C predicts [O{\small III}] more accurately than Model~D and this is not expected to be general behaviour. Nevertheless, the difference between the [O{\small III}] luminosity predicted by the fiducial model and Model~D is still significant as the luminosity in the fiducial model is 1.7$\times$ greater. We reiterate that the temperature used for all of these calculations is the non-equilibrium value from the simulation. In general, simulations run without the non-equilibrium processes will predict a different temperature so this 1.7$\times$ difference is likely to be a minimum. 

In summary, Model~D, in general, predicts the most accurate emission line luminosities. This is expected as it uses the most information from the simulation; however, the discrepancies are still large enough to warrant adopting a full non-equilibrium approach. These differences may be larger or smaller depending on environment (e.g., in the disc versus in an outflow) and we reserve this analysis for future work. Model~C, which is most similar to our previous work, predicts line luminosities with accuracy better than a factor of four for the lines considered. While not ideal, the approach can predict reasonable luminosities at significantly lower computational expense. However, we do not endorse the use of Models~A and B as the luminosity predictions are too inaccurate to make robust predictions of galaxy properties.

\section{Caveats}
\label{cavs}
Despite our improved ability to model individual metal ionisation states and the observable quantities that naturally emerge from such an approach, there remain numerous caveats to our existing work. First, we re-emphasise that certain, potentially important, physical processes are missing from our model. We have neglected Auger ionisation, dielectronic recombination suppression, and charge exchange with helium. Our model can be trivially extended to include these processes; however this comes with an extra computational cost. Thus the model, as presented, should only be used in situations where these processes are unimportant. The model may not be applicable in high density, self-shielded regions of the ISM where dust physics and cooling from molecules such as CO that are not currently included in {\small RAMSES-RTZ} may be important \citep[e.g.][]{Richings2014b,Klessen2016}. For additional computational efficiency, we have provided the functionality to artificially truncate the ionisation states of certain elements below their fully ionised values (e.g., for our example simulation we have only modelled six ionisation states of Mg and Si). We have seen in Figure~\ref{coll_ion_test_1} that this can cause the final two ionisation states to diverge from the expected equilibrium abundance for oxygen and potentially more for iron depending on the temperature. For this reason, we always recommend that if an ion is important for gas cooling or an observed quantity, the model should include at least two further ionisation states above the one of interest. Finally, in order to solve for level populations analytically, we have modelled gas cooling using only two-level or three-level ions. This is likely a fairly accurate approach at low temperatures but not the case at $T\gtrsim10^4$K. Extending our model to this regime is once again trivial; however, it comes with a significant extra computational cost which is why we have opted for the current approach of using tabulated ion-by-ion values from \cite{Oppenheimer2013}.

Like all simulations, finite spatial and mass resolution means that certain physical processes cannot be correctly modelled in the simulation. For example, the O{\small IV} ionised region is not fully resolved in our Stromgren sphere test. Most important for our current work is carefully choosing the spatial resolution and star particle mass so that the Stromgren radius is as well resolved as possible. If the Stromgren radius is not resolved, the simulation will likely miss some of the highly ionised states that exist close to the star particle which could bias the observable quantities measured from the simulation (see \citealt{Rosdahl2015b} for a similar discussion regarding radiation pressure). This may be why our 18pc resolution simulation falls below the SDSS galaxies on the BPT diagram as the O{\small III} abundance may be under-predicted. Therefore, modelling strong line diagnostics like the BPT diagram should be treated with special care. When the resolution is too poor, additional modelling may be needed to capture the physics very close to star forming regions as we have applied to our isolated disc simulation. The same argument holds if radiating black holes are included in the model. For this reason, our new code is particularly well suited for the CGM where, in principle, the physics can be more easily resolved compared to the ISM; however, even in this regime, finite resolution may cause issues \citep[e.g.][]{Peeples2019,Hummels2019,vdv2019}.

Depending on the SED, inaccuracies could arise due to the coarse sampling of radiation bins. We always recommend to use as many radiation energy bins as possible; however, since multi-frequency implementations of the M1 method for radiation transfer often scale with the number of frequency bins used, large numbers of bins are not always computationally efficient. We highlight the fact that it is not necessarily important that the bin edges occur at the ionisation energies of individual species. This was also demonstrated in our Stromgren sphere test. {\small RAMSES-RTZ} follows {\small RAMSES-RT} by using luminosity-weighted average cross sections for each species. For systems with low numbers of relatively similar bright sources (as is the case for the isolated disc), this approximation is reasonably accurate. However, if a simulation includes sources with very different SEDs (e.g., AGN and stars), for greater accuracy, either more energy bins should be used or cross sections for each cell should be calculated based on only local sources.

We also highlight the fact that our model is limited by the accuracy of the reaction rates that are available. In many cases, rates have only been determined over a specific temperature range and must be extrapolated in some way to be applicable to all astrophysical environments probed by our simulation. In certain cases, reaction rates are unavailable (e.g., many He charge exchange reactions) which can cause biases in the model. Finally, not all reaction rates are agreed upon in the literature (e.g., the hydrogen recombination rates used by {\small RAMSES-RT} \citep{Hui1997} are different from those used in {\small CLOUDY}). It is beyond the scope of this work to reconcile these differences, rather we stress that for important reactions, the choice of reaction rate can lead to systematic biases when comparing models.

When including subgrid models for unresolved physics that impact metal ionisation states such as those that return mass into the gas-phase (e.g., SN feedback), there is no obvious choice for how to update the ionisation states upon the return of mass. In this work, we have chosen not to update the ionisation states of the metals upon mass return, which is equivalent to assuming that the ionisation states in the returned material are consistent with the surroundings; however, other methods could be used. For example, the material could be returned in a completely neutral or a completely ionised state. In many cases, the ionisation states will rapidly update to reflect the local gas and radiation conditions; however, caution should be used when analysing simulation outputs that occur very shortly after one of these events because it is imperative that the simulation has had a chance to evolve the reaction network before conclusions can be made about the physical properties of the system. 

Finally, we emphasise that the work in the previous section is not presented to demonstrate that the specific galaxy formation model used in this work is able to reproduce observations. The contrived initial conditions and lack of cosmological environment inhibit a detailed comparison. Rather, our primary goal is to demonstrate the ease in which observable quantities can be extracted from the simulations and the importance on non-equilibrium effects.

\section{Conclusions}
\label{dac}
We have presented {\small RAMSES-RTZ}, an implementation of non-equilibrium metal chemistry and cooling in the {\small RAMSES-RT} code. The code is designed to model the ionisation states and cooling from an arbitrary number of metal ions including physics such as photoionisation, collisional ionisation, charge exchange, radiative recombination, and dielectronic recombination. Our model follows the ionisation states of C, N, O, Mg, Si, S, Fe and Ne as well as cooling at $T\leq10^4$K from O{\small I}, O{\small II}, O{\small III}, C{\small I}, C{\small II}, Si{\small I}, Si{\small II}, N{\small II}, Fe{\small I}, Fe{\small II}, S{\small I}, and Ne{\small II}.

While we are not the first to implement non-equilibrium metal chemistry and cooling into a simulation, other codes either follow a limited number of ionisation states \citep[e.g.][]{Baczynski2015,Lupi2020}, are not cosmological \citep[e.g][]{Ziegler2018}, are not coupled with on-the-fly radiation hydrodynamics \citep[e.g.][]{Oppenheimer2013,Richings2014}, or employ RT methods that are not well-suited to large cosmological simulations \citep[e.g.][]{Sarkar2021}. For these reasons, we have developed an implementation in the {\small RAMSES-RT} code that addresses the latter three issues. Furthermore, we have provided interfaces to two different ODE solvers to address the expense that comes with solving large networks as we find that our fast solver maintains the accuracy that is achieved with an implicit BDF solver but can result in speed-ups of $\sim15\times$. The main limitation is the memory cost of following so many ionisation states. For example, following all ions of the metals in our fiducial model would increase the number of cell properties from $\sim18$ to more than 100 (i.e. $>5\times$) which can lead to a substantial increase in communication time between processors for MPI runs. Therefore, artificially truncating the maximum or minimum ionisation state could be an enticing solution depending on the problem of interest.

The code is suitable in a variety of astrophysical environments including the CGM and galactic outflows where the ionisation states may be out of equilibrium, as well as in the optically thin ISM where we have explicitly coupled the cooling to the metal ionisation states via the emission of various fine-structure lines. Furthermore, the code was specifically designed to model various observable quantities, in particular, emission and absorption lines that require knowledge of the ionisation states of each metal as well as their collisional partners.

We have presented various benchmarks that demonstrated a good agreement with {\small CLOUDY} when equilibrium has been reached for each of the major physical processes (collisional ionisation, photoionisation, and charge exchange) that impact metal ionisation states. Furthermore, we demonstrated that if a particular ion is of interest, an accurate model requires following at least two further ionised states if the atom is artificially truncated in a non-fully ionised configuration in order to accurately model the system.  

\section*{Acknowledgements}
HK thanks the anonymous referee for their comments that greatly improved the manuscript. HK thanks Joki Rosdahl, Taysun Kimm, and Martin Rey for their support and insight throughout this project as well as Romain Teyssier for making {\small RAMSES} publicly available.

\section*{Data Availability}
The data underlying this article will be shared on reasonable request to the corresponding author.

\bibliographystyle{mnras}
\bibliography{example}




\bsp	
\label{lastpage}
\end{document}